\documentclass{ws-acs}

\usepackage[usenames,dvipsnames]{color}
\usepackage{amssymb}
\usepackage{subfigure}
\usepackage{graphicx}
\usepackage{xspace}
\usepackage{lipsum}
\usepackage{url}

\begin{document}


\DeclareGraphicsExtensions{.eps,.pdf,.png,.jpg,.jpeg}

\newcommand{\jbullet}[0]{{\itshape jbullet}\xspace}
\newcommand{\colt}[0]{{\itshape colt}\xspace}
\newcommand{\jung}[0]{{\itshape jung}\xspace}
\newcommand{\lucene}[0]{{\itshape lucene}\xspace}
\newcommand{\internet}[0]{{\itshape internet}\xspace}
\newcommand{\social}[0]{{\itshape collaboration}\xspace}

\newcommand{\exmplwidth}[0]{0.25\textwidth}
\newcommand{\groupswidth}[0]{0.18\textwidth}
\newcommand{\figswidth}[0]{0.725\textwidth}
\newcommand{\subfigswidth}[0]{0.50\textwidth}
\newcommand{\plotswidth}[0]{0.90\textwidth}
\newcommand{\logplotswidth}[0]{1.00\textwidth}

\newcommand{\define}[1]{{\itshape #1}\xspace}

\newcommand{\secref}[1]{Section~\ref{sec:#1}\xspace}
\newcommand{\defref}[1]{Definition~\ref{def:#1}\xspace}
\newcommand{\equref}[1]{Eq.~(\ref{equ:#1})\xspace}
\newcommand{\figref}[1]{Figure~\ref{fig:#1}\xspace}
\newcommand{\tblref}[1]{Table~\ref{tbl:#1}\xspace}

\newcommand{\page}[1]{{\color{RawSienna}\scriptsize\textbf{#1 PAGE(S)}}\xspace}
\newcommand{\todo}[0]{{\color{RawSienna}\large\textbf{TODO}}\xspace}
\newcommand{\redo}[1]{{\color{RawSienna}\textit{#1}}\xspace}
\newcommand{\refs}[0]{{\color{RawSienna}\textbf{[REF]}}\xspace}
\newcommand{\cits}[0]{{\color{RawSienna}\textbf{[CIT]}}\xspace}

\makeatletter
\def\@biblabel#1{[#1]}
\makeatother


\markboth{L. \v{S}ubelj et al.}{Node mixing and group structure of complex software networks}
\catchline{}{}{}{}{}

\title{NODE MIXING AND GROUP STRUCTURE OF\\ COMPLEX SOFTWARE NETWORKS}

\author{LOVRO \v{S}UBELJ}
\address{University of Ljubljana, Faculty of Computer and Information Science,\\
Tr\v{z}a\v{s}ka cesta 25, SI-1001 Ljubljana, Slovenia\\
lovro.subelj@fri.uni-lj.si}

\author{SLAVKO \v{Z}ITNIK}
\address{University of Ljubljana, Faculty of Computer and Information Science,\\
Tr\v{z}a\v{s}ka cesta 25, SI-1001 Ljubljana, Slovenia\\
slavko.zitnik@fri.uni-lj.si}

\author{NELI BLAGUS}
\address{University of Ljubljana, Faculty of Computer and Information Science,\\
Tr\v{z}a\v{s}ka cesta 25, SI-1001 Ljubljana, Slovenia\\
neli.blagus@fri.uni-lj.si}

\author{MARKO BAJEC}
\address{University of Ljubljana, Faculty of Computer and Information Science,\\
Tr\v{z}a\v{s}ka cesta 25, SI-1001 Ljubljana, Slovenia\\
marko.bajec@fri.uni-lj.si}

\maketitle

\begin{history}
\received{(received date)}
\revised{(revised date)}
\accepted{(day month year)}
\comby{(xxxxxxxxxx)}
\end{history}


\begin{abstract}
Large software projects are among most sophisticated human-made systems consisting of a network of interdependent parts. Past studies of software systems from the perspective of complex networks have already led to notable discoveries with different applications. Nevertheless, our comprehension of the structure of software networks remains to be only partial. We here investigate correlations or mixing between linked nodes and show that software networks reveal dichotomous node degree mixing similar to that recently observed in biological networks. We further show that software networks also reveal characteristic clustering profiles and mixing. Hence, node mixing in software networks significantly differs from that in, e.g., the Internet or social networks. We explain the observed mixing through the presence of groups of nodes with common linking pattern. More precisely, besides densely linked groups known as communities, software networks also consist of disconnected groups denoted modules, core/periphery structures and other. Moreover, groups coincide with the intrinsic properties of the underlying software projects, which promotes practical applications in software engineering.
\end{abstract}

\keywords{Software networks; node mixing; node groups; software engineering.}


\section{\label{sec:introduction}Introduction}
Large software projects are one of the most sophisticated and diverse human-made systems; still, our comprehension of their complex structure and behavior remains to be only partial~\cite{CY09}. On the other hand, studies on modeling software systems as networks of interdependent parts have recently led to some notable discoveries and promoted different applications~\cite{FBL11,SB11s}. Complex networks possibly provide the most adequate framework for the analysis of large software systems developed according to object-oriented, structured programming and other paradigms~\cite{Mye03,TCMT13}.

Past studies have already shown that software systems modeled as directed networks are \define{scale-free}~\cite{BA99} with a power-law in-degree distribution and, e.g., exponential out-degree distribution~\cite{VCS02,VS05a}. Furthermore, networks are \define{small-world}~\cite{WS98}, when represented with undirected graphs~\cite{Mye03,Koh09}, and reveal a hierarchical~\cite{VS07} and fractal structure~\cite{CLMPT06,BSB12}. The latter can be, similarly as the properties mentioned above, related to code complexity or reusability and the quality of the underlying software projects~\cite{SB12s,TCMT13}. Authors have also proposed different growing models of software networks~\cite{VCS02,SFMV03,LZCXA13} and investigated the importance of particular nodes in the networks~\cite{Koh09}, their evolution during project execution~\cite{CY09}, practical applications of network community and motif structure~\cite{VS05b,SB11s}, and other~\cite{SB12s}. 

In the present paper, we first analyze the correlations or \define{mixing}~\cite{New02,New03a} between linked nodes in software networks, which has not yet been addressed properly. Despite a common belief that software networks are negatively correlated or \define{disassortative} by degree~\cite{New03a,GXYLG10} as, e.g., web graphs or the Internet~\cite{PVV01}, we show that networks are indeed strongly disassortative by in-degree, but much more positively correlated or \define{assortative} by out-degree, otherwise a characteristic property of different social networks~\cite{NP03}. Software networks thus reveal \define{dichotomous} degree mixing, similar to that recently detected in undirected biological networks~\cite{HL11}.

\begin{figure}[!t]
	\centerline{\psfig{file=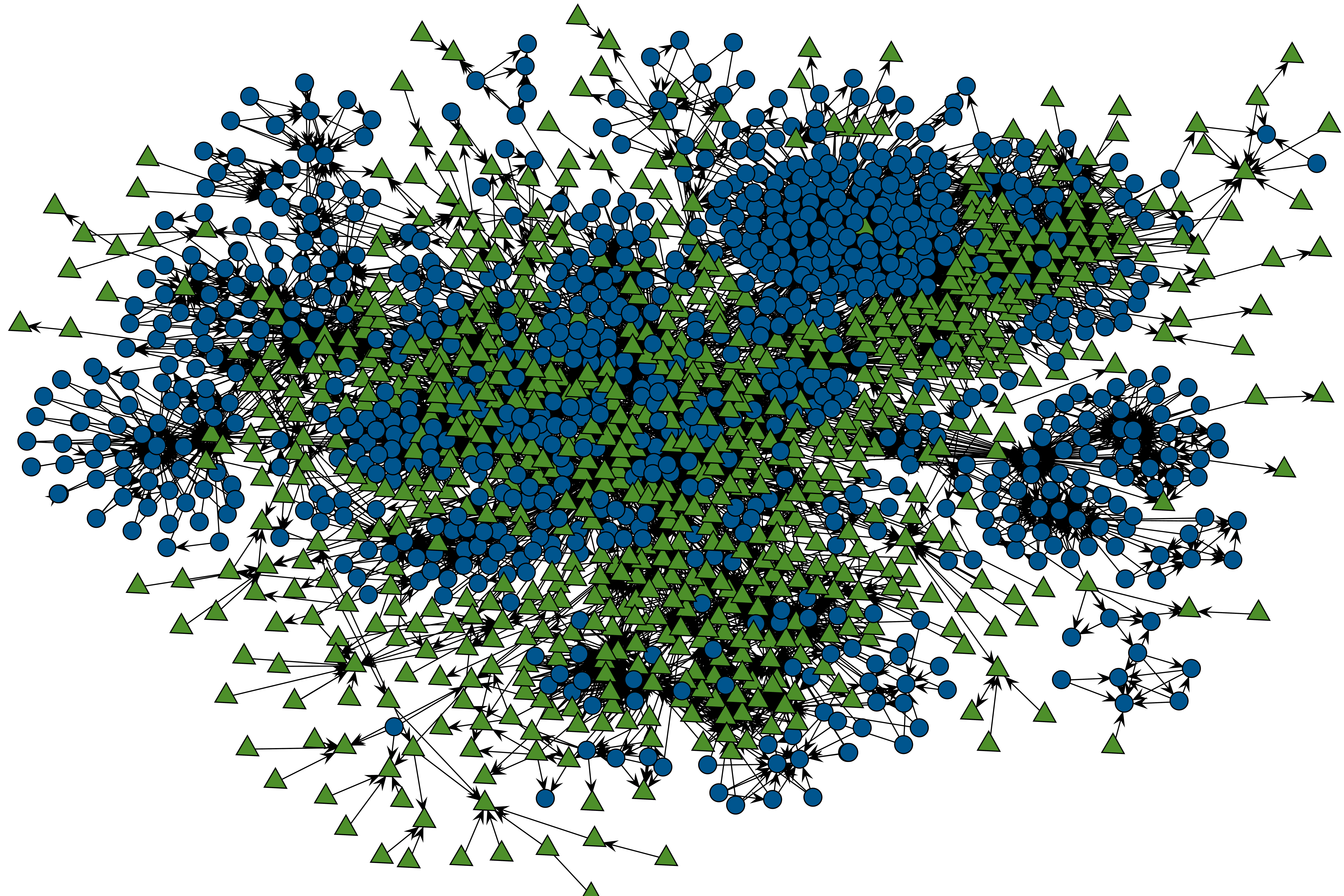, width=\figswidth}}
	\caption{\label{fig:lucene}Software dependency network representing the Lucene search engine. (Nodes with degree-corrected clustering~\cite{SV05} above or below the mean are shown as circles and triangles, respectively.)}
\end{figure}

We further show that software networks are characterized by a sickle-shaped \define{clustering}~\cite{WS98} profile also observed in~\cite{HL11}. This unique shape is retained in the case of \define{degree-corrected clustering}~\cite{SV05}, whereas the structure of the networks differs significantly from that of the Internet or a social network. More precisely, software networks contain connected parts or regions with very low or very high degree-corrected clustering (\figref{lucene}), which is else observed only for either the Internet or a social network. Nevertheless, all types of networks reveal clear degree-corrected clustering assortativity that has not been reported in the literature before.

We explain the observed degree mixing and clustering assortativity through the presence of different types of \define{groups} or \define{clusters} of nodes with common linking pattern~\cite{NL07}. Besides densely linked groups denoted \define{communities}~\cite{GN02}, software networks also consist of groups of structurally equivalent nodes denoted \define{modules}~\cite{SB12u}, and different \define{mixtures}~\cite{SBB13} of these, with core/periphery and hub \& spokes structures as special cases. We stress that the existence of different types of groups implies high clustering assortativity, with sparse module-like groups occupying regions with very low clustering and dense community-like groups in regions with higher clustering. While the former explain the observed disassortativity by degree, the latter in fact promote the assortativity in the out-degree. Note that the conclusions are consistent with the results obtained for the Internet and a social network, where mostly module-like or community-like groups are found, respectively.

Although the main purpose of the analysis of node mixing is to relate characteristic group structure to the existing network properties, the dichotomous degree mixing in fact implies many of the common properties of real-world networks~\cite{HL11} (e.g., robustness). The latter, together with the observed node clustering assortativity, might be of independent interest in network model design and other.

The paper does not provide a clear rationale behind the existence of different types of groups in software networks. Nevertheless, the revealed groups are found to closely coincide with some of the intrinsic properties of the underlying software projects. The paper thus also includes preliminary work and results of selected applications of network group detection in software engineering.

The rest of the paper is structured as follows. For the analysis in the paper, we adopt software dependency networks based on~\cite{SB11s,SB12s}, which are introduced in~\secref{networks}. Next, \secref{mixing} contains an extensive empirical analysis and formal discussion on node degree and clustering mixing. Analysis of the characteristic groups of nodes in software networks is conducted in~\secref{groups}, while some practical applications of group detection in software engineering are given in~\secref{applications}. \secref{conclusions} concludes the paper and gives prominent directions for future work.

\section{\label{sec:networks}Software dependency networks}
Complex software systems can be modeled with various types of networks including software architecture maps~\cite{VCS02}, class diagrams~\cite{VS05b}, inter-package~\cite{LW06} and class dependency networks~\cite{SB11s}, class, method and package collaboration graphs~\cite{HCK06}, software mirror~\cite{CY09} and subroutine call graphs~\cite{Mye03}, to name just a few. Networks mainly divide whether they are constructed from source code, byte code or program execution traces, and due to the level of software architecture represented by the nodes and the types of software relationships represented by the links.

\begin{figure}[!t]
\centerline{
\subfigure[\label{fig:example:class}]{
\psfig{file=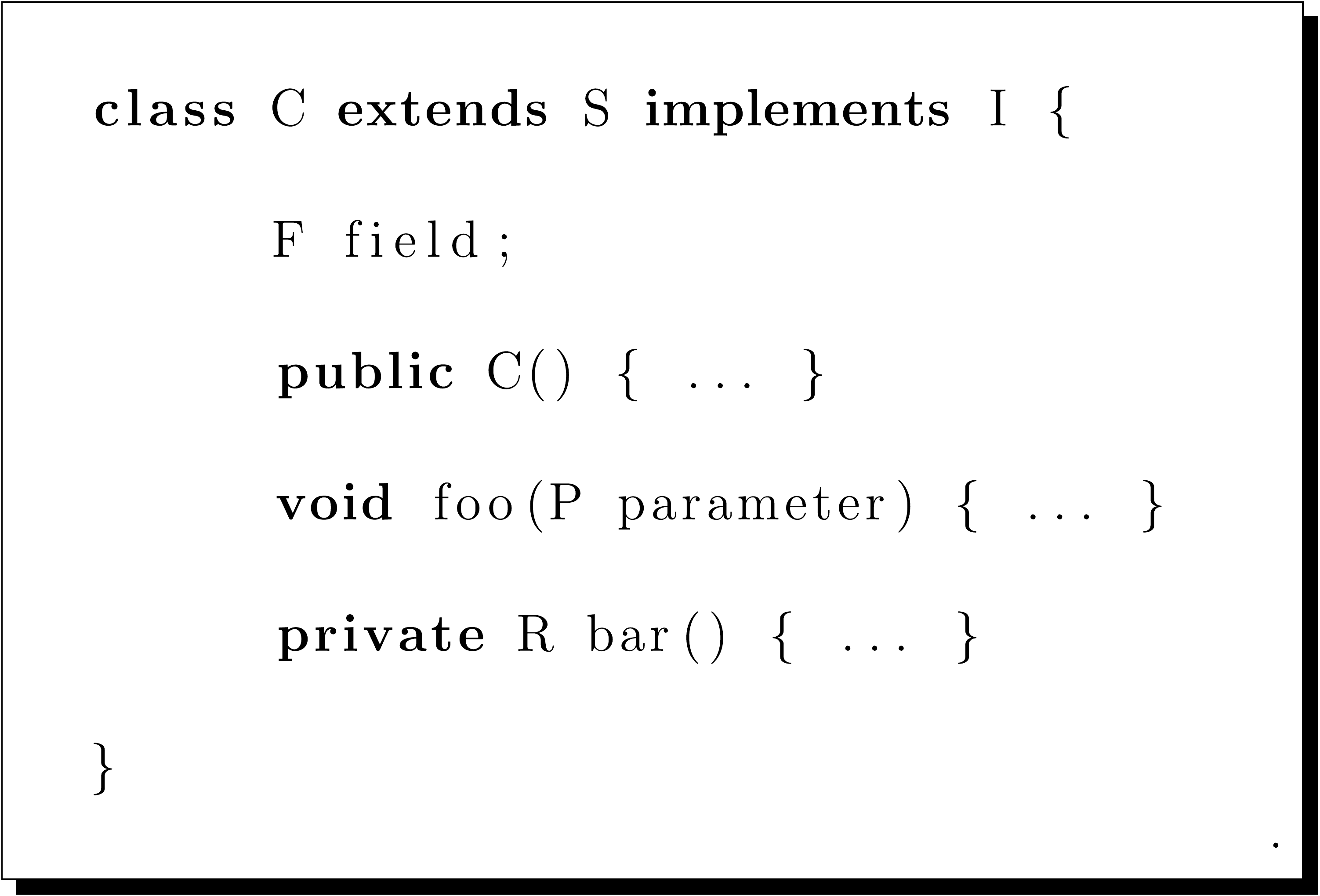, width=\exmplwidth}}
\subfigure[\label{fig:example:networks}]{
\psfig{file=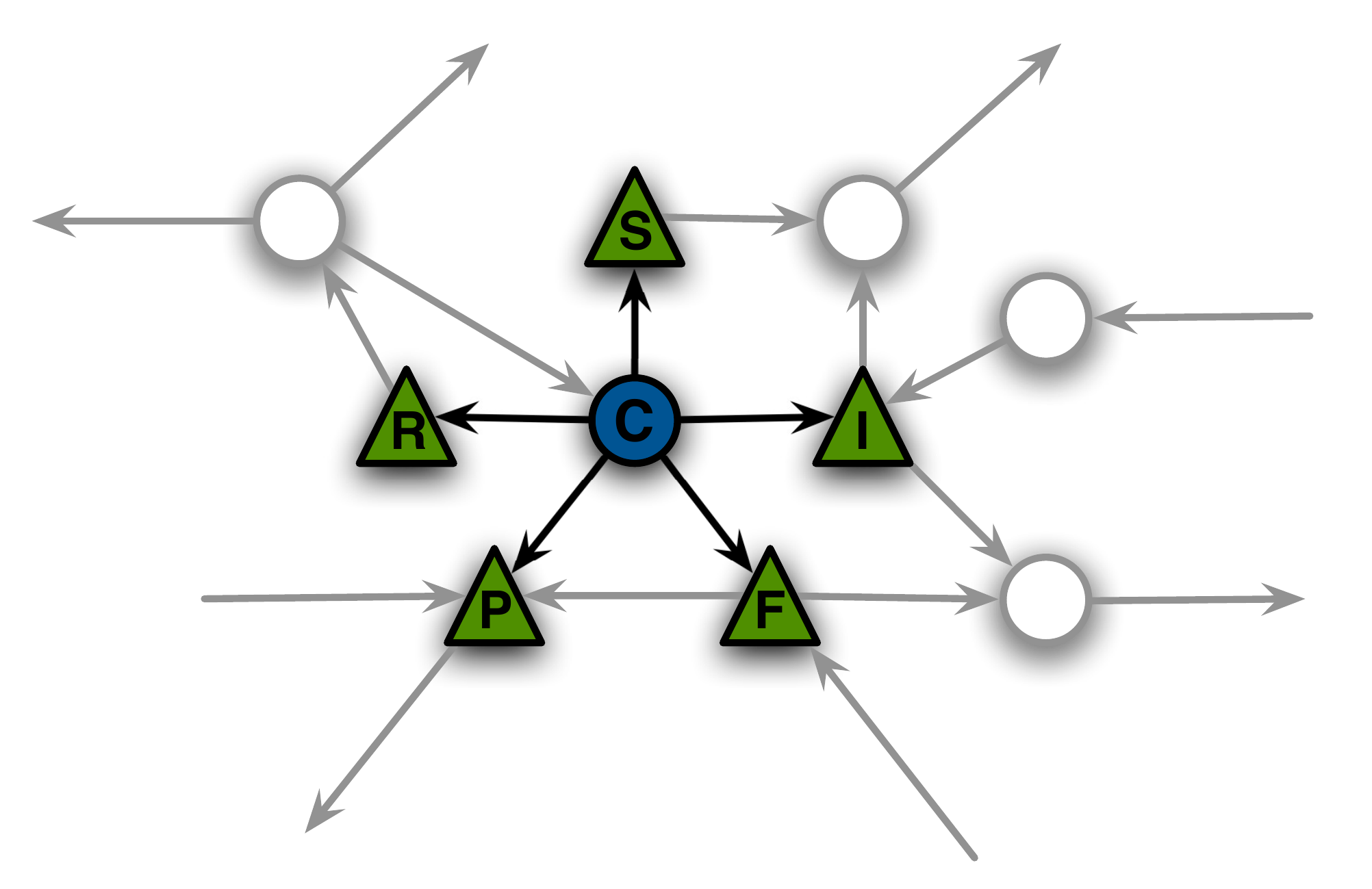, width=0.30\textwidth}}}
	\caption{\label{fig:example}(a) A toy example class written in Java and (b) the corresponding class dependency network.}
\end{figure}

For consistency with most past work, we consider class dependency networks~\cite{SB11s,SB12s} that are suitable for modeling object-oriented software systems. Here, nodes represent software classes and links correspond to different types of dependencies among them (e.g., inheritance). More formally, let a software project consist of classes $C=\left\{C_1,C_2,\dots\right\}$. Corresponding class dependency network is a directed graph $G(V,L)$, where $V=\left\{1,2,\dots,n\right\}$ is the set of nodes and $L$ is the set of links, $m=|L|$. Class $C_i$ is represented by a node $i\in V$, while a directed link $\left(i,j\right)\in L$ corresponds to some dependency between classes $C_i$ and $C_j$ (\figref{example}). This can be either an \textit{inheritance} (i.e., $C_i$ extends class or implements interface $C_j$), a \textit{composition} (i.e., $C_i$ contains a field or variable of type $C_j$) or a \textit{dependence} (i.e., $C_i$ contains a constructor, method or function with parameter or return type $C_j$).

Note that class dependency networks are constructed merely from the signatures of software classes, and fields and functions therein. Thus, the networks address the inter-class structure of the software systems, whereas the intra-class dependencies are ignored~\cite{SB12s}. However, as such information is often decided by a team of developers, prior to the actual software development, it is not influenced by the programming style of each individual developer. Moreover, such networks coincide with the flow of information and also the human comprehension of object-oriented software systems. Nevertheless, the networks still give only a partial view of the~system.

According to the object-oriented programming paradigm, a class that extends a parent class also inherits all of its functionality (not considering the visibility). Hence, each class implicitly acquires the dependencies of its parent class, the parent class of its parent class, and so on. For the analysis in the paper, we thus first construct the networks based on the explicit class dependencies as described above, while we then copy also the implicit dependencies of each class from its parent classes. This provides somewhat more adequate representation of the intrinsic structure of the software system and also coincides with the developer's view. Note that the process does not significantly increase the overall number of dependencies (see below). Finally, we reduce the networks to simple directed graphs, to limit the influence of individual developers as above. Networks thus utilize merely the connectedness between the nodes, while disregarding its strength. We consider four such software dependency networks that are shown in~\tblref{networks} (see also~\figref{lucene}). All selected networks represent well-known software projects developed in Java including physics simulation, scientific computing and network analysis libraries.

\begin{table}[!h]
\tbl{\label{tbl:networks}Software, Internet and social networks used in the study. (The values in brackets show the number of links corresponding to explicit class dependencies.)}
{\begin{tabular}{@{}cccc@{}} \toprule
Network & Description & $n$ & $m$ \\ \colrule
\jbullet & JBullet 2.72 game physics simulation toolbox & $\hphantom{0}166$ & $\hphantom{0}619$ \hphantom{0}($552$) \\
\colt & Colt 1.2.0 scientific \& technical computing library & $\hphantom{0}227$ & $\hphantom{0}963$ \hphantom{0}($709$) \\
\jung & JUNG 2.0.1 network \& graph analysis framework & $\hphantom{0}306$ & $\hphantom{0}930$ \hphantom{0}($713$) \\
\lucene & Lucene 4.1.0 high-performance text search engine & $1657$ & $6808$ ($6252$) \\ \colrule
\internet & Oregon 2003 autonomous systems snapshot~\cite{LKF05} & $\hphantom{0}767$ & $1857$ \hphantom{ (0}-\hphantom{00)} \\
\social & Network scientists collaborations~\cite{New06b} & $1589$ & $2742$ \hphantom{ (0}-\hphantom{00)} \\ \botrule
\end{tabular}}
\begin{tabnote}
Software networks are reduced to largest connected components
\end{tabnote}
\end{table}

For a thorough empirical comparison in the following sections, we also consider two other real-world networks. Namely, a snapshot of communications between autonomous systems of the Internet collected by the University of Oregon in 2003~\cite{LKF05} and a social network of collaborations between scientists working on network theory and experiment~\cite{New06b} (\tblref{networks}). These are simple undirected networks. Although some directed social and technological networks would enable more straightforward comparison, such networks are commonly either much larger than software networks or do not reveal particularly clear group structure. On the other hand, we stress that the selected networks represent two fundamentally different topologies. While social networks are characterized by a dense degree assortative structure and community-like groups~\cite{New02,NP03}, the Internet is much sparser and disassortative by degree~\cite{PVV01}. Also, the prevalent groups of nodes are module-like, e.g., hub \& spokes~\cite{LKSF10}.

\section{\label{sec:mixing}Node mixing in software networks}
The present section contains an extensive comparative analysis of different networks according to node degree and clustering mixing. We first review characteristics of node degree distributions in~\secref{mixing:scale-free} and then show that software networks reveal dichotomous degree mixing in~\secref{mixing:dichotomous}. Next, sickle-shaped clustering profiles of software networks are explored in~\secref{mixing:sickle}, while \secref{mixing:clustering} provides empirical evidence of node clustering assortativity in real-world networks.

\subsection{\label{sec:mixing:scale-free}Scale-free node degree distributions}
Let $k_i$ be the degree of node $i\in V$ and let $\left<k\right>$ be the mean degree in the network. 
For directed networks, the degree is defined as the sum of in-degree and out-degree.
Next, let $\Delta$ be the maximum degree, and $\Delta_{in}$ and $\Delta_{out}$ the maximum in-degree and out-degree, respectively. Last, let $\gamma$ be the scale-free exponent of the power-law degree distribution $P(k)\sim k^{-\gamma}$~\cite{BA99}, $\gamma>1$, and let $\gamma_{in}$ and $\gamma_{out}$ be the exponents corresponding to in-degree and out-degree distributions, respectively. The values of $\gamma$-s were estimated by maximum-likelihood method with goodness-of-fit tests~\cite{CSN09}.

\begin{figure}[!t]
	\centerline{\psfig{file=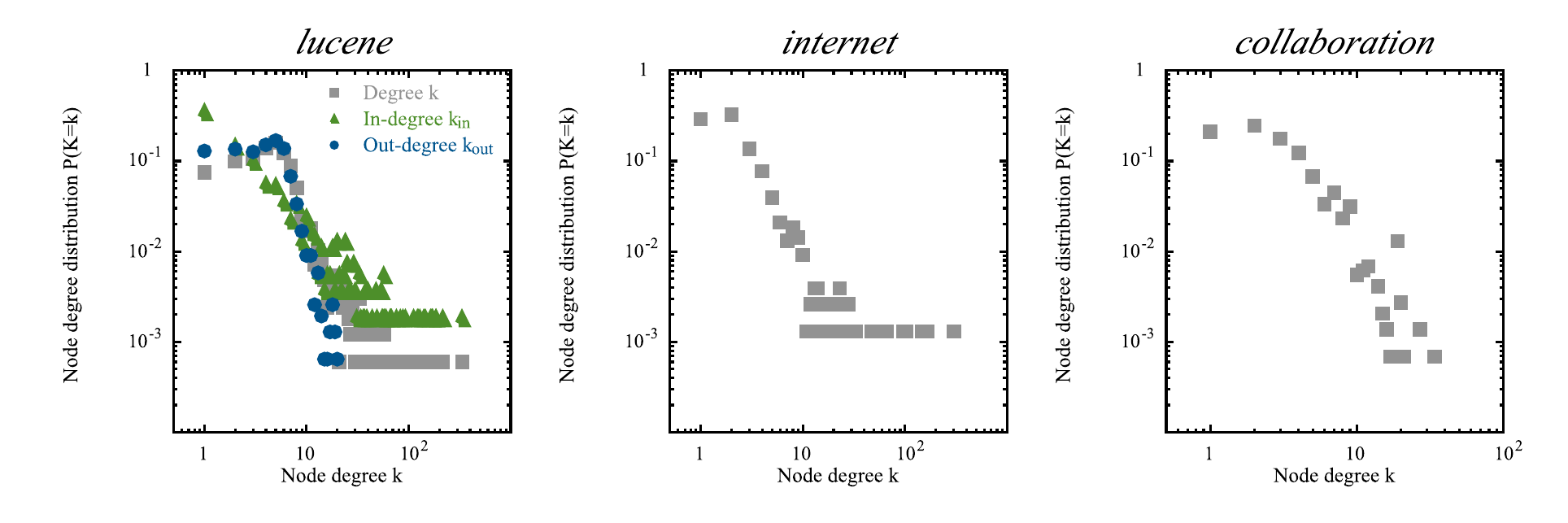, width=\logplotswidth}}
	\caption{\label{fig:mixing:scale-free}Node degree distributions of larger networks (see also \tblref{mixing:scale-free}). Note that \lucene software network reveals scale-free (in-)degree distribution as the Internet and a truncated, e.g., log-normal or exponential, out-degree distribution more similar to the \social network.}
\end{figure}

\tblref{mixing:scale-free} describes node degree sequences of different networks. The degree $\left<k\right>$ is somewhat comparable across software networks and approximately half the size for \internet and \social networks. Observe, however, that in the case of directed software networks the values of $\Delta$-s and $\gamma$-s are obviously governed by a much broader in-degree sequences, compared to a relatively suppressed out-degree sequences (e.g., \lucene network). Particularly, as past work has already shown, software networks have scale-free in-degree distribution that follows a power-law with $2<\gamma_{in}<3$~\cite{VCS02} and highly truncated, e.g., log-normal~\cite{CMPS07} or exponential~\cite{VS05a}, out-degree distribution (see~\tblref{mixing:scale-free}). Note also that the tail of the (in-)degree distribution of \lucene software network is well modeled by the scale-free degree distribution of a sparse topology of the Internet, while, from the perspective of out-degrees, the network is somewhat more similar to a dense assortative social network (\figref{mixing:scale-free}).

\begin{table}[!h]
	\tbl{\label{tbl:mixing:scale-free}Node degree sequences of different networks. (The exponents $\gamma$-s in italics do not represent a valid fit to a power-law~\cite{CSN09}.)}
	{\begin{tabular}{@{}cccccccc@{}} \toprule
		Network & $\left<k\right>$ & $\Delta$ & $\Delta_{in}$ & $\Delta_{out}$ & $\gamma$ & $\gamma_{in}$ & $\gamma_{out}$ \\ \colrule
		\jbullet & $7.46$ & $\hphantom{0}62$ & $\hphantom{0}62$ & $22$ & $2.80$ & $2.26$ & ${\it 4.04}$ \\
		\colt & $8.48$ & $140$ & $140$ & $13$ & $2.56$ & $2.56$ & ${\it 3.91}$ \\
		\jung & $6.08$ & $\hphantom{0}95$ & $\hphantom{0}92$ & $12$ & $2.65$ & $2.77$ & ${\it 4.47}$ \\
		\lucene & $8.22$ & $337$ & $333$ & $20$ & $2.24$ & $2.14$ & ${\it 4.91}$ \\ \colrule
		\internet & $4.68$ & $303$ & - & - & $2.28$ & - & - \\
		\social & $3.45$ & $\hphantom{0}34$ & - & - & $2.85$ & - & - \\ \botrule
	\end{tabular}}
\end{table}

For the concerned software dependency networks, in-degree and out-degree sequences have a rather clear meaning in software engineering. The out-degree of node $i$ corresponds to the number of classes required to implement the functionality of class $C_i$ and is thus a measure of  'external' complexity~\cite{SB12s}. Indeed, different software quality metrics are based on the out-degrees of nodes in software networks~\cite{CK94,TCMT13}. On the other hand, the in-degree of node $i$ corresponds to the number of classes that depend on or use class $C_i$ and is related to the level of code reusability~\cite{SB12s}.

Highly reused classes are, obviously, well known among developers and are thus also more commonly used in the future. The latter is exactly the principle behind the preferential attachment model~\cite{BA99}, which produces power-law in-degree distribution in software dependency networks~\cite{SB12s}. For the case of the out-degree distribution, long scale-free tail is suppressed by constant incremental refactoring of classes within a growing software project~\cite{BFNRSVMT06} (to reduce its complexity), while such distribution also results from a certain class of software duplication mechanisms~\cite{VS05a}.

\subsection{\label{sec:mixing:dichotomous}Dichotomous node degree mixing}
The most straightforward way to analyze node degree mixing in general networks is to measure $r$~\cite{New02,New03a}, which is defined as a Pearson correlation coefficient of degrees at links' ends, $r\in\left[-1,1\right]$. Hence,
\begin{equation}
r=\frac{1}{2\sigma_k}\sum_{(i,j)\in L}\left(k_i-\left<k\right>\right)\left(k_j-\left<k\right>\right),
\label{equ:r}
\end{equation}
where $\sigma_k$ is the standard deviation, i.e., $\sigma_k=\sqrt{\sum_{i\in V}\left(k_i-\left<k\right>\right)^2}$. Assortative mixing by degree shows as a positive correlation $r>0$, while disassortative degree mixing refers to a negative correlation $r<0$. For the case of directed networks, one can similarly define four additional coefficients $r_{(\alpha,\beta)}$~\cite{FFGP10}, $\alpha,\beta\in\left\{in,out\right\}$, where $\alpha$, $\beta$ correspond to the types of degrees of links' source and target nodes,~respectively.

\begin{table}[!b]
\tbl{\label{tbl:mixing:dichotomous}Node degree mixing coefficients~\cite{GXYLG10} of different networks.}
{\begin{tabular}{@{}cccccc@{}} \toprule
Network & $r$ & $r_{(in,in)}$ & $r_{(in,out)}$ & $r_{(out,in)}$ & $r_{(out,out)}$ \\ \colrule
\jbullet & $-0.21$ & $-0.29$ & $-0.07$ & $-0.26$ & $-0.14$ \\
\colt & $-0.24$ & $-0.27$ & $-0.06$ & $-0.25$ & $-0.28$ \\
\jung & $-0.22$ & $-0.25$ & $-0.05$ & $-0.24$ & $-0.13$ \\
\lucene & $-0.28$ & $-0.30$ & $\hphantom{-}0.00$ & $-0.29$ & $-0.04$ \\ \colrule
\internet & $-0.26$ & - & - & - & - \\
\social & $\hphantom{-}0.46$ & - & - & - & - \\ \botrule
\end{tabular}}
\end{table}

\tblref{mixing:dichotomous} summarizes degree mixing in different networks. As already stated before, 
social networks reveal strong assortative mixing~\cite{New02} (e.g., \social network), whereas the Internet is degree disassortative~\cite{PVV01}. Software networks also appear to be disassortative by degree according to $r$~\cite{GXYLG10}. Nevertheless, this is actually a consequence of the prevailing in-degree sequences (see~\secref{mixing:scale-free}). The networks are indeed highly disassortative by in-degree, $r_{(in,in)}\ll 0$, though much more assortative by out-degree in most cases, $r_{(out,out)}\gg r_{(in,in)}$ (e.g., \lucene network). Expectedly, $r_{(in,out)}$ reveals no clear mixing regime, $r_{(in,out)}\approx 0$, while $r_{(out,in)}$ is again governed by the dominant in-degrees, $r_{(out,in)}\approx r_{(in,in)}$.

\begin{figure}[!t]
	\centerline{\psfig{file=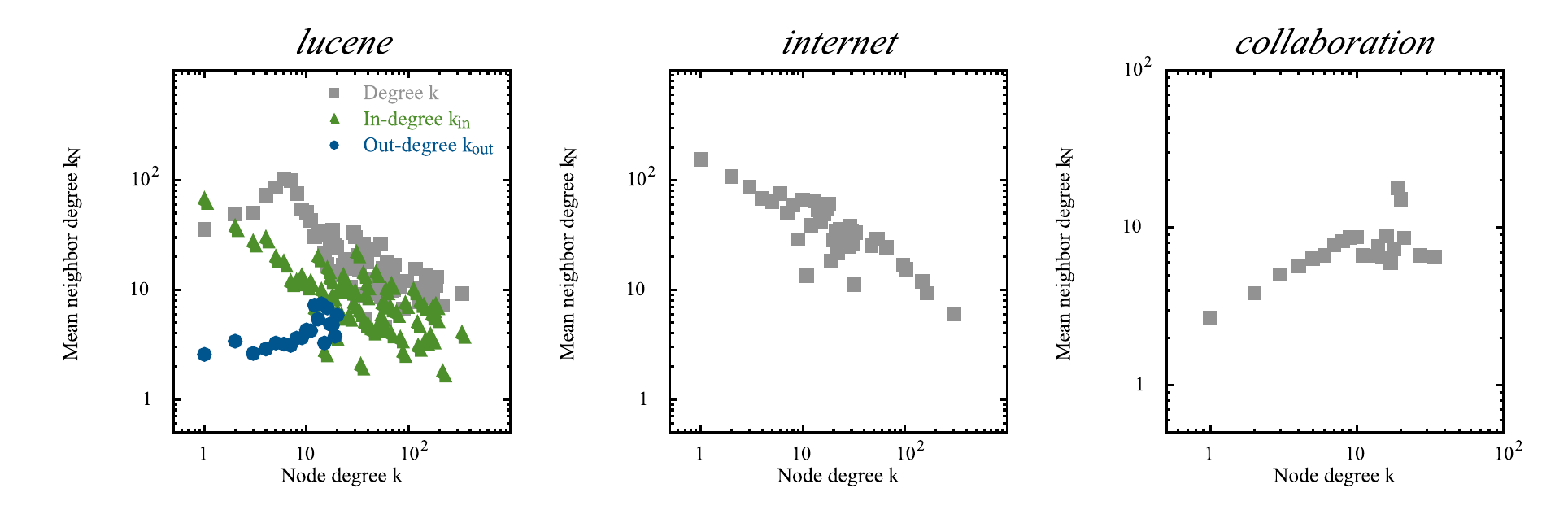, width=\logplotswidth}}
	\caption{\label{fig:mixing:dichotomous}Neighbor connectivity plots~\cite{PVV01} of larger networks (see also \tblref{mixing:dichotomous}). Note that \lucene software network reveals dichtomous degree mixing that is disassortative by in-degree as the Internet and assortative by out-degree as social networks  (e.g., \social network).}
\end{figure}

Note that above coefficients provide a rather limited global view of degree mixing and can capture merely linear correlations. \figref{mixing:dichotomous} shows also neighbor connectivity plots~\cite{PVV01} that display mean neighbor degree $k_N$ against node degree $k$. Here, assortative or disassortative mixing reflects in either increasing or decreasing trend, respectively. While the software network is clearly disassortative by in-degree, it is in fact slightly assortative by out-degree, as in the case of a social network. Furthermore, the degrees $k$ show a clear two-phase or dichotomous mixing that is controlled by out-degrees for smaller $k$, and by in-degrees, when $k$ increases. Although one can also observe some dichotomous behavior for \social and \internet networks, this does not appear significant and can be due to the size of the networks. Thus, as previously claimed, software networks reveal dichotomous degree mixing and differ from other degree disassortative networks like web graphs~and~the~Internet. 

It ought to be mentioned that similar observations were recently made also in undirected biological networks~\cite{HL11}. Although these are disassortative by degree~\cite{MS02}, removing a certain percentage of high degree nodes or \define{hubs}~\cite{HBHGBZDWCRV04} renders the networks degree assortative. Since hubs in software networks correspond to nodes with high in-degree (see~\tblref{mixing:scale-free}), our work generalizes that in~\cite{HL11} to directed networks.

Dichotomous degree mixing in software networks can be seen as a product of different programming paradigms. Recall that the out-degree of a node measures the complexity of the corresponding software class, whereas its in-degree is related to class reuse (see~\secref{mixing:scale-free}). Disassortativity in the in-degrees can be interpreted as low probability of hubs to link; thus, highly reused classes tend not to depend on each other. Since these commonly implement a rather different functionality, the latter is in fact a result of minimum-coupling and maximum-cohesion principle~\cite{SMC99}. On the other hand, object-oriented software systems are commonly developed according to Lego hypothesis~\cite{BFNRSVMT06}, where smaller and simpler classes are used to implement larger and more complex ones, and so on.  As this results in an entire hierarchy of classes with increasing complexity across the levels of the hierarchy, a class depends only on classes with rather similar complexity, i.e., classes from the previous level. Obviously, this implies assortativity in the out-degrees in software networks.

\subsection{\label{sec:mixing:sickle}Sickle-shaped node clustering profiles}
Besides degree distributions and mixing considered above, real-world networks are commonly assessed due to their transitivity. For simple undirected graphs, this can be measured by node clustering coefficient $c$~\cite{WS98}, $c\in\left[0,1\right]$, defined as
\begin{equation}
c_i=\frac{t_i}{{k_i \choose 2}},
\label{equ:c}
\end{equation}
where $t_i$ is the number of links between the neighbors of node $i\in V$ and ${k_i \choose 2}$ is the maximal number of links ($c_i=0$ for $k_i\leq 1$). Note that the denominator in~\equref{c} introduces biases in the definition, since ${k_i \choose 2}$ often cannot be reached due to a fixed degree sequence~\cite{SV05} (see below). Thus, an alternative definition of node degree-corrected clustering coefficient $d$~\cite{SV05}, $d\in\left[0,1\right]$, has been proposed as
\begin{equation}
d_i=\frac{t_i}{\omega_i},
\label{equ:d}
\end{equation}
where $\omega_i$ is the maximal possible number of links between the neighbors of node $i$ with respect to their degrees ($d_i=0$ for $k_i\leq 1$). Since $\omega\leq {k \choose 2}$, $d\geq c$ by definition.

\begin{table}[!b]
\tbl{\label{tbl:mixing:sickle}Node clustering coefficients of different networks.}
{\begin{tabular}{@{}ccccc@{}} \toprule
Network & ~~~$\left<c\right>$~~~ & ~~~$\left<d\right>$~~~ & ~~~$d=1$~~~ & ~~~$d<p$~~~ \\
 & & & \multicolumn{2}{c}{(\% nodes)} \\ \colrule
\jbullet & $0.43$ & $0.50$ & $\hphantom{0}9\%$ & $20\%$ \\
\colt & $0.50$ & $0.58$ & $17\%$ & $13\%$ \\
\jung & $0.51$ & $0.58$ & $19\%$ & $19\%$ \\
\lucene & $0.50$ & $0.55$ & $11\%$ & $13\%$ \\ \colrule
\internet &$0.29$ & $0.32$ & $21\%$ & $55\%$ \\
\social & $0.64$ & $0.69$ & $61\%$ & $28\%$ \\ \botrule
\end{tabular}}
\begin{tabnote}
Networks are reduced to simple undirected~graphs
\end{tabnote}
\end{table}

\tblref{mixing:sickle} shows the mean node (degree-corrected) clustering $\left<c\right>$ and $\left<d\right>$ in different networks. As these are small-world~\cite{WS98}, $\left<c\right>$ and $\left<d\right>$ are considerably larger than the expected clustering coefficient $p$ in a corresponding random graph~\cite{ER59}, $p=\left<k\right>/\left(n-1\right)$. The structure of \social network else reveals the most densely linked neighborhoods, where the majority of nodes have $d$ equal to one (see~\tblref{mixing:sickle}). Exactly the opposite holds for \internet network, where $d$ is close to zero, $d<p$, in most cases. On the other hand, software networks are again characterized by an interplay between the dense structure of social networks and the sparse topology of the Internet. Most of the nodes have moderate values of $d$, $p<d<1$, whereas nodes with either very low or high $d$ are concentrated in certain parts of the networks (not shown).

\begin{figure}[!t]
	\centerline{\psfig{file=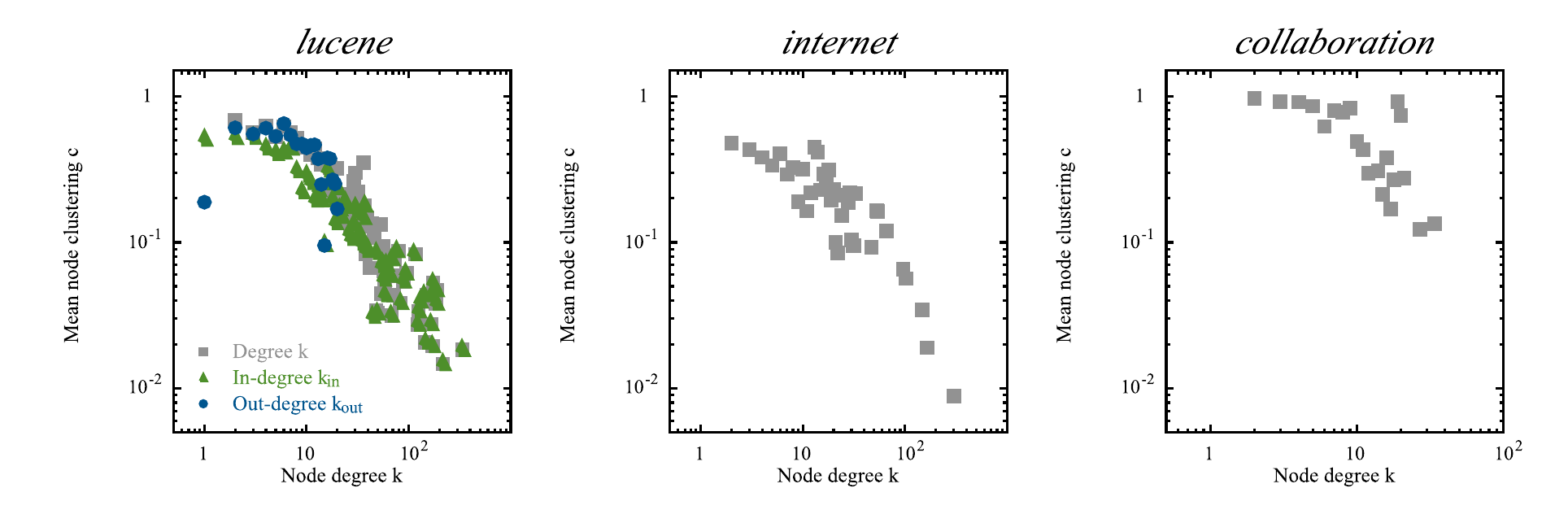, width=\logplotswidth}}
	\caption{\label{fig:mixing:sickle:c}Node clustering~\cite{WS98} profiles of larger networks (see also~\tblref{mixing:sickle}). Note degree biases introduced in the standard definition of clustering that imply low values for hubs, which is particularly apparent in degree disassortative networks (e.g., \lucene and \internet networks).}
\end{figure}

\begin{figure}[!t]
	\centerline{\psfig{file=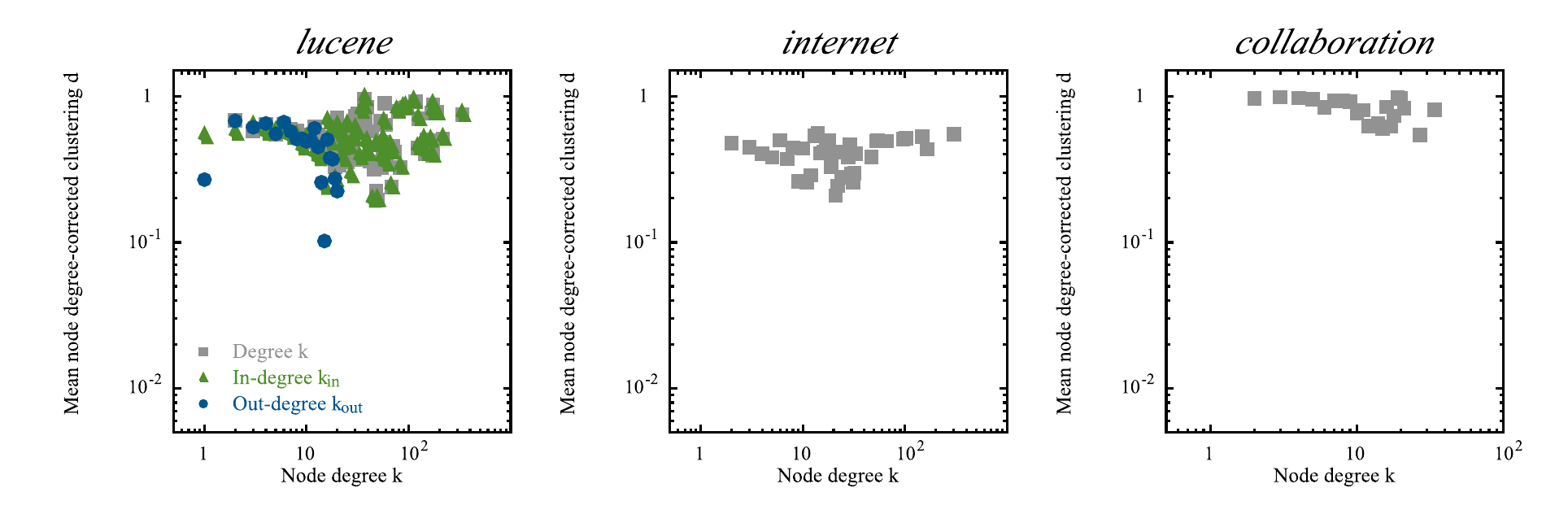, width=\logplotswidth}}
	\caption{\label{fig:mixing:sickle:d}Node degree-corrected clustering~\cite{SV05} profiles of larger networks (see also~\tblref{mixing:sickle}). Note that \lucene software network reveals a sickle-shaped clustering profile most notably pronounced for out-degrees, which is absent in the case of the Internet and the \social network.}
\end{figure}

We next consider node (degree-corrected) clustering profiles shown in \figref{mixing:sickle:c} and \figref{mixing:sickle:d}. One can observe degree biases in the standard definition of clustering $c$ that imply low $c$ for hubs (see~\equref{c}), particularly apparent in degree disassortative networks (see~\figref{mixing:sickle:c}). More precisely, $c$ decreases rapidly with $k$, roughly following a power-law form $c\sim k^{-1}$ in the case of the Internet~\cite{VPV02,SV05}. Note that these biases are absent from the degree-corrected definition of clustering $d$ (see~\figref{mixing:sickle:d}), which thus provides somewhat more adequate measure of network transitivity.

Notice also very peculiar sickle-shaped (degree-corrected) clustering profiles revealed for the software network (see, e.g., \figref{mixing:sickle:c}). This unique form is most notably pronounced in the case of out-degrees and is, at least in the undirected case, an artifact of dichotomous node degree mixing~\cite{HL11}. On the contrary, profiles of \internet and \social networks show no particular scaling for degree-corrected clustering $d$ (see \figref{mixing:sickle:d}), consistent with the analysis of node degree mixing in~\secref{mixing:dichotomous}. Nevertheless, all networks considered here reveal clear degree-corrected clustering assortativity, which is throughly investigated in the following section.

Same as before, (degree-corrected) clustering profiles in software networks can be related to the intrinsic properties of the underlying software systems~\cite{SB11s,SB12s}. While nodes that represent core classes of a software project commonly group together into dense neighborhoods with high clustering, nodes with lower clustering most often correspond to different implementations of the same functionality (see~\figref{applications}).

\subsection{\label{sec:mixing:clustering}Node degree-corrected clustering assortativity}
\begin{figure}[!t]
	\centerline{\psfig{file=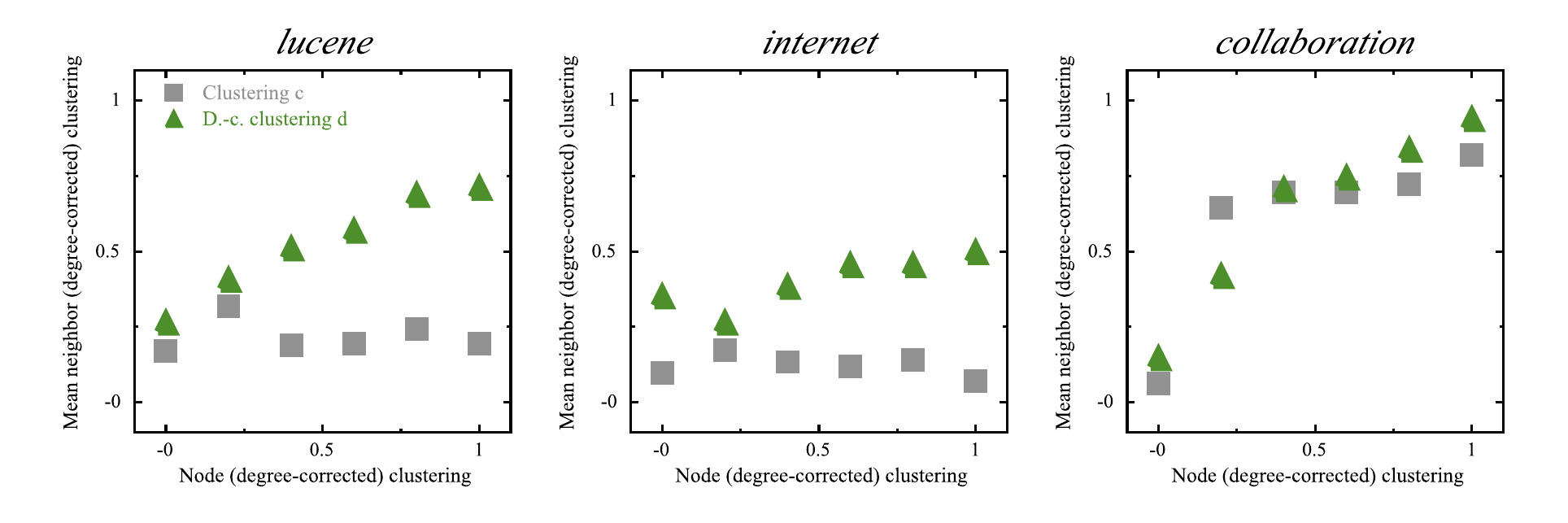, width=\plotswidth}}
	\caption{\label{fig:mixing:clustering}Neighbor (degree-corrected) clustering plots of larger networks (see also~\tblref{mixing:clustering}). Note~that all networks reveal a clear degree-corrected clustering~\cite{SV05} assortativity (e.g., \lucene network), which is absent from the standard definition of clustering~\cite{WS98} (e.g., \internet network).}
\end{figure}

The present section explores node (degree-corrected) clustering mixing in different networks. For this purpose, we define clustering mixing coefficient $r_c$, $r_c\in\left[-1,1\right]$,~as
\begin{equation}
r_c=\frac{1}{2\sigma_c}\sum_{(i,j)\in L}\left(c_i-\left<c\right>\right)\left(c_j-\left<c\right>\right)
\label{equ:rc}
\end{equation}
and similarly $r_d$ for degree-corrected clustering coefficient. $r_c$ and $r_d$ are again just Pearson correlation coefficients of (degree-corrected) clustering at links' ends and are shown in~\tblref{mixing:clustering}. Due to degree biases in $c$ (see~\secref{mixing:sickle}), $r_c>0$ in degree assortative networks (e.g., \social network), while $r_c<0$ for networks that are disassortative by degree (e.g., \lucene network). On the other hand, all networks show clear degree-corrected clustering assortativity with $r_d\gg 0$ (see also \figref{mixing:clustering}). Note also that correlations reflected in $r_d$ are much stronger than in the case of degree mixing coefficients $r$-s (see~\tblref{mixing:dichotomous}). To the best of our knowledge, this distinctive property of real-world networks has not yet been reported in the literature.

\begin{table}[!h]
\tbl{\label{tbl:mixing:clustering}Node clustering mixing coefficients of different networks.}
{\begin{tabular}{@{}ccc@{}} \toprule
Network & ~~~$r_c$~~~ & ~~~$r_d$~~~ \\
 & & \\ \colrule
\jbullet &  $-0.06$ & $0.50$ \\
\colt & $-0.26$ & $0.35$ \\
\jung & $-0.07$ & $0.33$ \\
\lucene & $-0.40$ & $0.50$ \\ \colrule
\internet & $-0.23$ & $0.26$ \\
\social & $\hphantom{-}0.44$ & $0.68$ \\ \botrule
\end{tabular}}
\begin{tabnote}
Networks are reduced to simple undirected graphs
\end{tabnote}
\end{table}

According to \secref{mixing:sickle}, nodes in software networks have very different values of degree-corrected clustering $d$, which is not true for social networks or the Internet. Together with strong  assortativity $r_d\gg 0$, this in fact implies entire connected parts or regions of nodes with rather similar $d$ (e.g., very low or high). The latter can be clearly seen in~\figref{lucene}, while, in the following section, we explain degree-corrected clustering assortativity, and dichotomous degree mixing observed in~\secref{mixing:dichotomous}, through the presence of characteristic groups of nodes with common linking pattern~\cite{NL07}. More precisely, dense community-like groups occupy network regions with higher $d$ and imply degree assortativity, while sparse module-like groups are found in regions with lower $d$ and are responsible for degree disassortativity.

\section{\label{sec:groups}Group structure of software networks}
Node group structure of different networks is explored using a principled group extraction framework based on~\cite{SBB13,ZLZ11a}. The present section thus first introduces the framework and corresponding formalisms in~\secref{groups:extraction}, while \secref{groups:structure} reports the characteristic group structure revealed in software and other networks. Last, \secref{groups:mixing} relates different types of groups to degree and clustering mixing observed in~\secref{mixing}, which uniquely characterizes the structure of these networks.

\subsection{\label{sec:groups:extraction}Node group extraction framework}
\begin{figure}[!t]
\centerline{\subfigure[\label{fig:framework:community}Community]{
\psfig{file=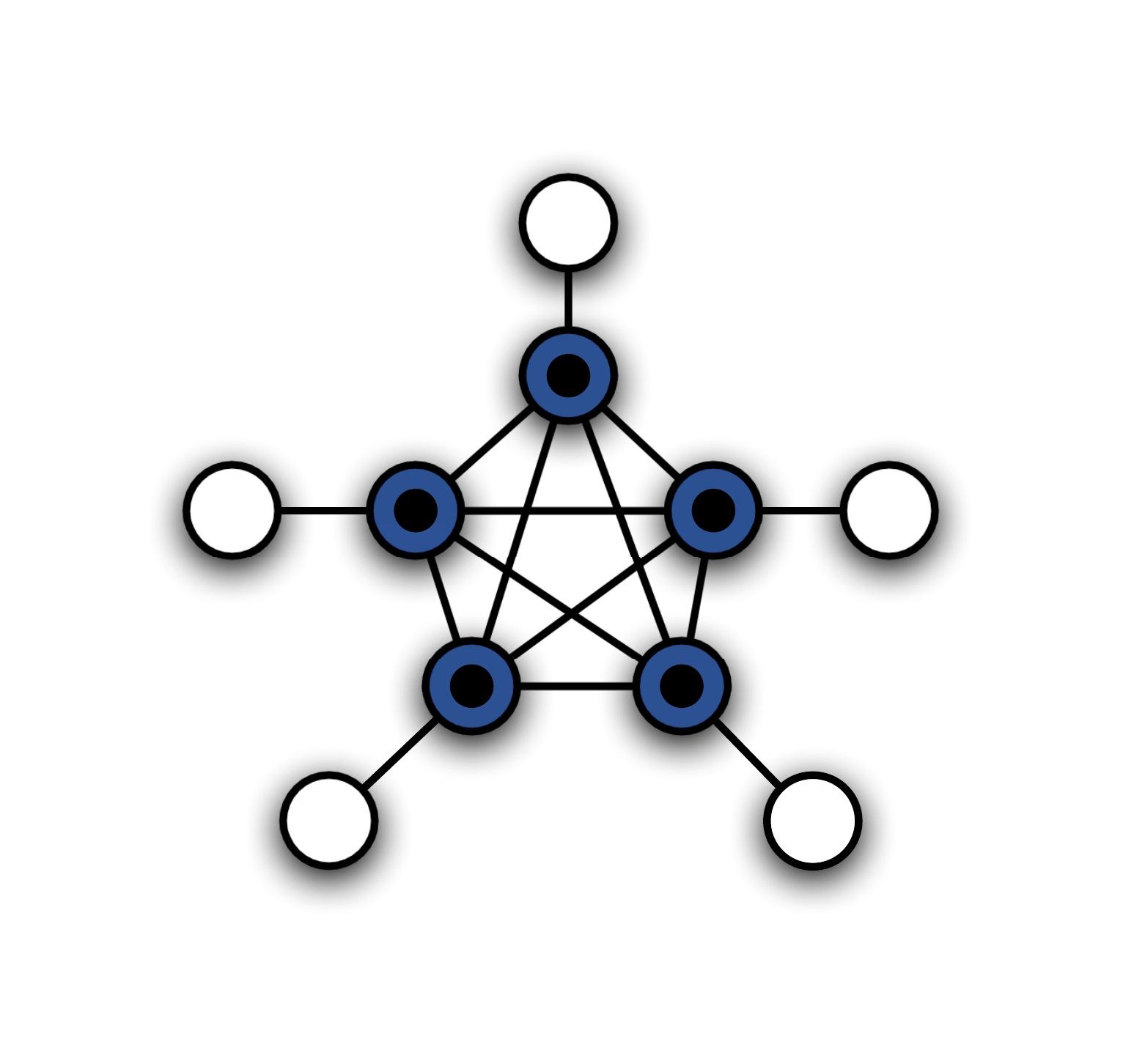, width=\groupswidth}}
\subfigure[\label{fig:framework:core}Core/periph.]{
\psfig{file=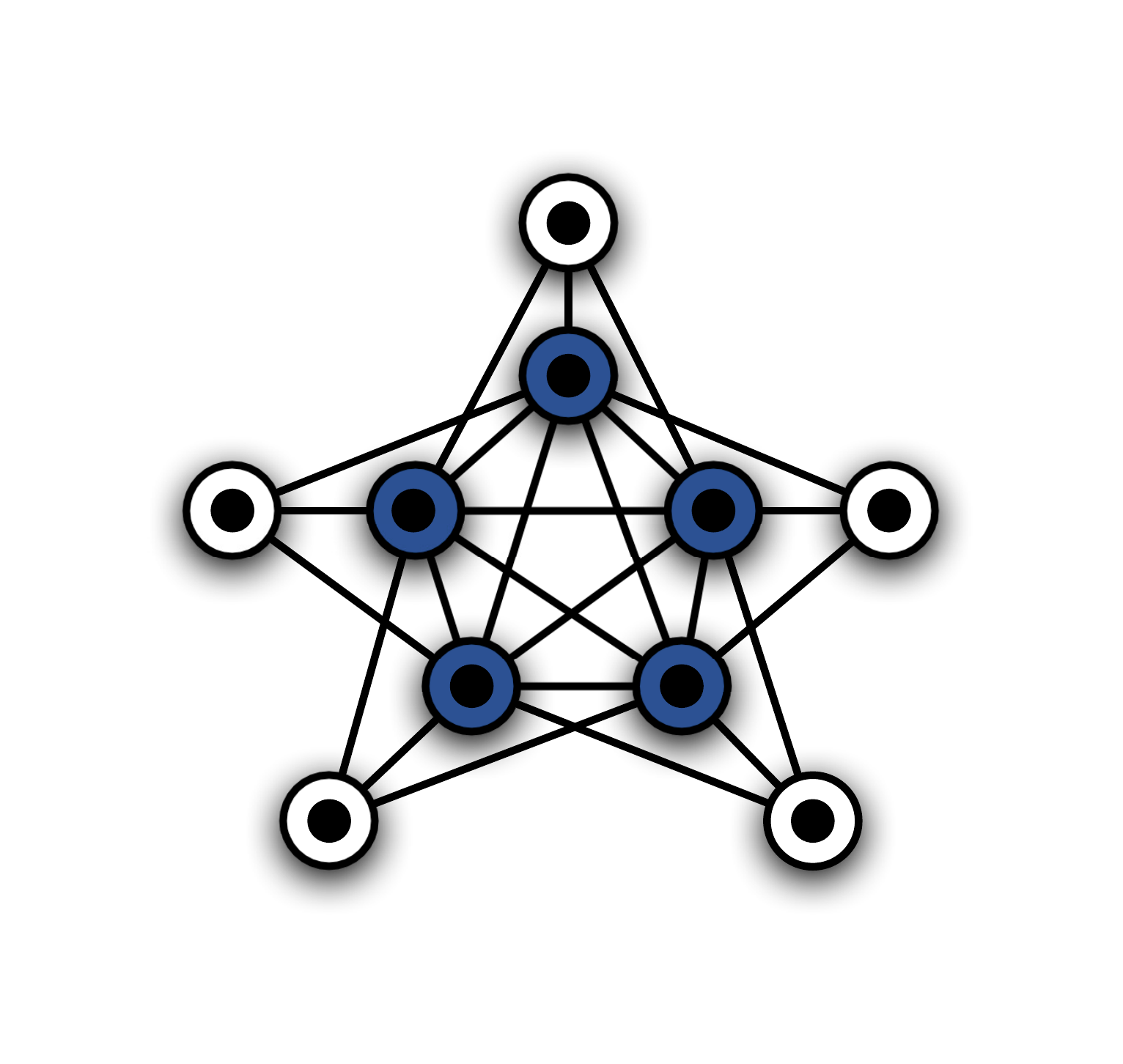, width=\groupswidth}}
\subfigure[\label{fig:framework:mixture}Mixture]{
\psfig{file=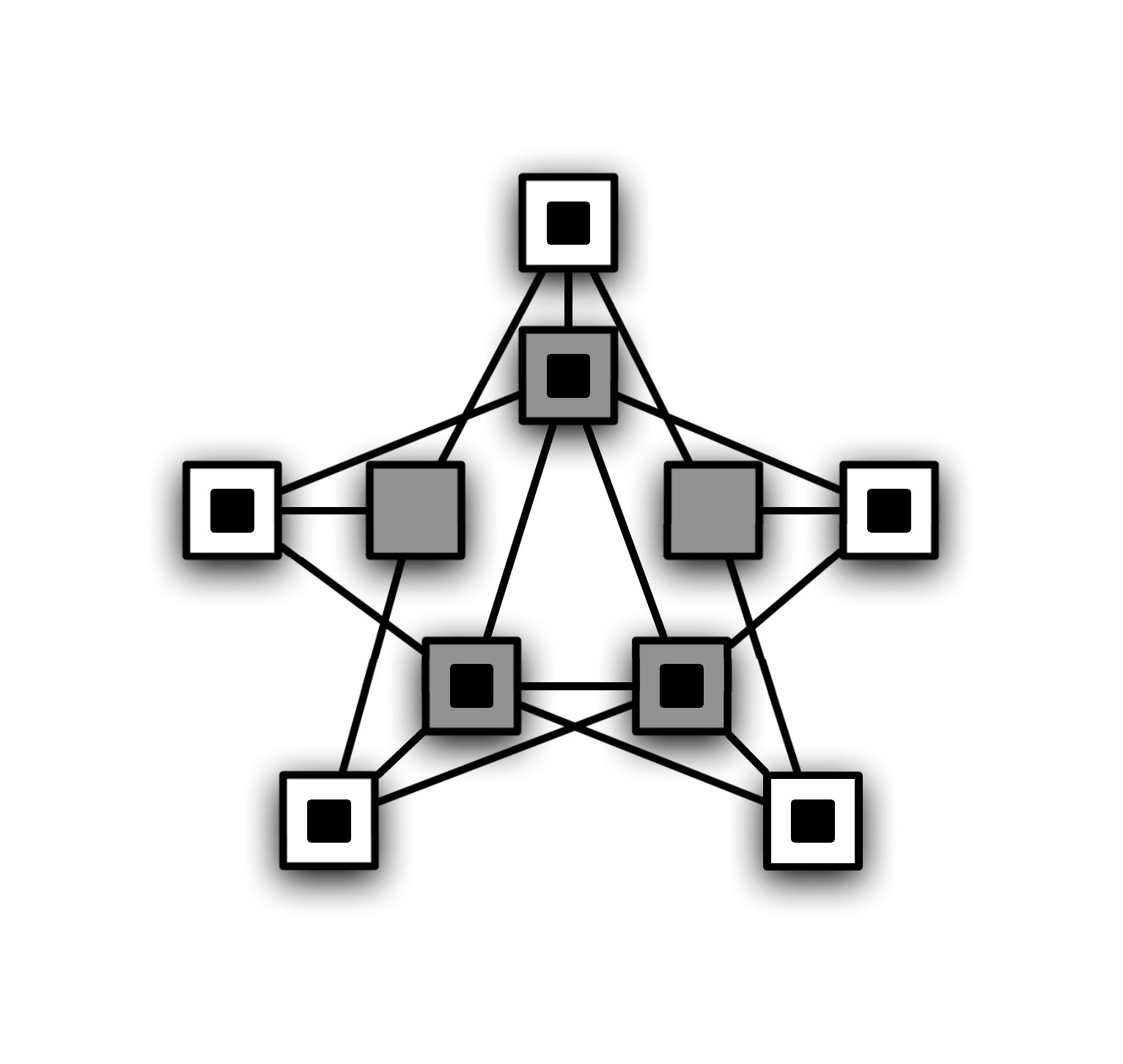, width=\groupswidth}}
\subfigure[\label{fig:framework:module}Module]{
\psfig{file=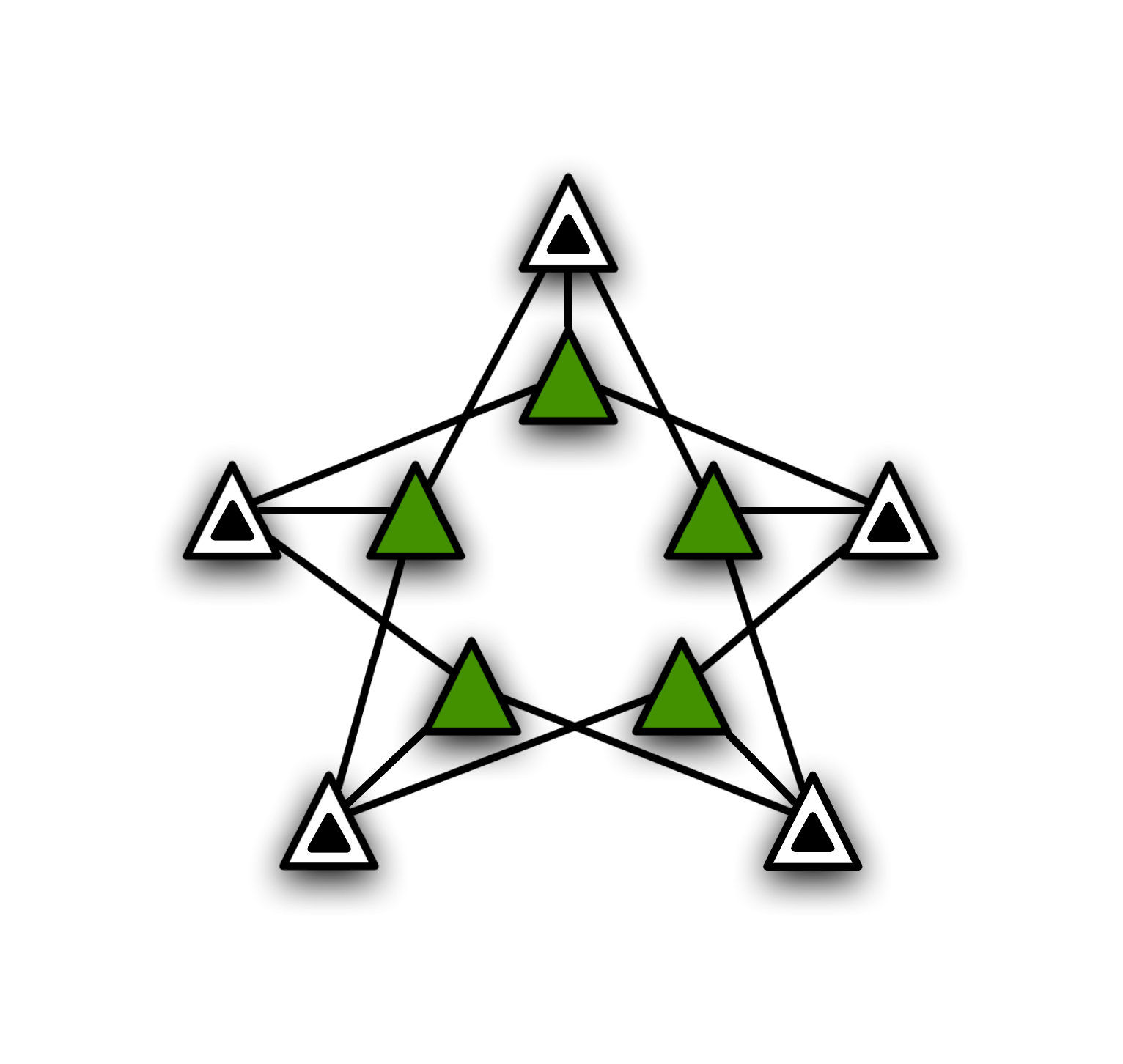, width=\groupswidth}}
\subfigure[\label{fig:framework:spokes}Hub \& spokes]{
\psfig{file=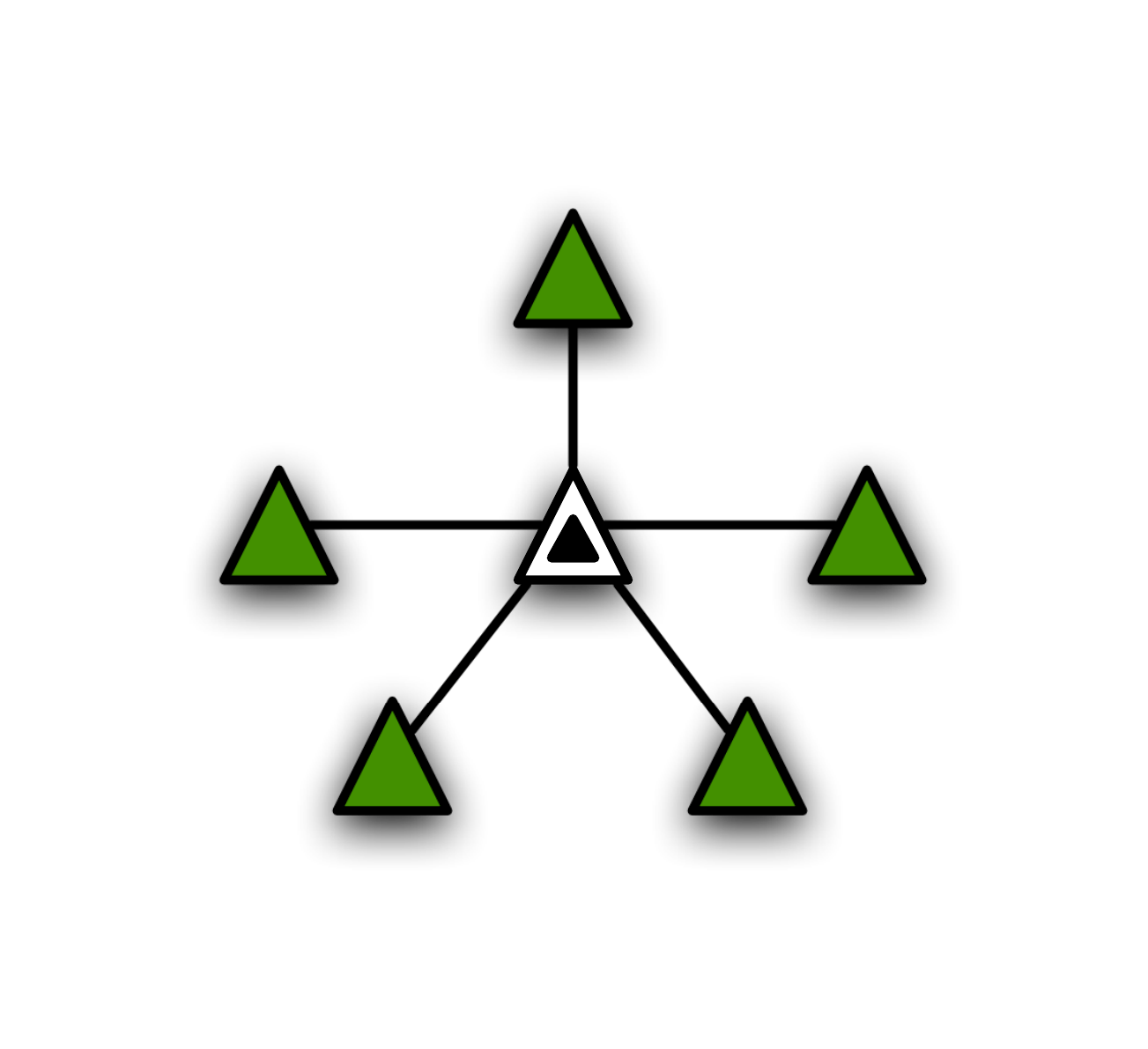, width=\groupswidth}}}
	\caption{\label{fig:framework}Toy examples of different types of groups of nodes in real-world networks (see also text). (Groups $S$ and corresponding patterns $T$ are shown with filled and marked nodes, respectively.)}
\end{figure}

The formalism proposed in~\cite{SBB13} defines network groups for the case of simple undirected graphs. Let $S$ be a group of nodes and $T$ a subset of nodes representing its characteristic linking pattern, $S,T\subseteq V$. Also, let $s=|S|$ and $t=|T|$. The node pattern $T$ is defined thus to maximize the number of links between $S$ and $T$, and minimize the number of links between $S$ and $T^C$, while disregarding the links with both endpoints in $S^C$. Note that this simple formalism allows one to derive most types of groups commonly analyzed in the literature~\cite{For10,New12} (\figref{framework}).

For instance, communities~\cite{GN02}, i.e., densely linked groups of nodes that are only sparsely linked between, are characterized by $S=T$. On the other hand, $S\cap T=\emptyset$ corresponds to groups of structurally equivalent~\cite{LW71} nodes denoted modules~\cite{SB12u}. Communities and modules represent two extreme cases, with all other groups being the mixtures of the two~\cite{SBB13}. For the analysis in the paper, we thus distinguish between three types of groups according to the following definitions.

\begin{definition}
\label{def:community}
{\itshape Community} is a group of nodes $S$ with $S=T$.
\end{definition}

\begin{definition}
\label{def:module}
{\itshape Module} is a group of nodes $S$ with $S\cap T=\emptyset$.
\end{definition}

\begin{definition}
\label{def:mixture}
{\itshape Mixture} is a group of nodes $S$ with $S\cap T\subset S,T$.
\end{definition}

All these groups have been extensively analyzed in the past~\cite{Sch07,POM09,For10,New12}. Clear communities appear in different social and information networks~\cite{GN02,DDDA05}, while modules are most commonly found in the case the Internet, biological and technological networks~\cite{PSR10,SB12u}. For consistency, we also consider two special cases.

\begin{definition}
\label{def:core-periphery}
{\itshape Core/periphery} structure is a mixture $S$ with either $S\subset T$ or $T\subset S$.
\end{definition}

\begin{definition}
\label{def:hub-spokes}
{\itshape Hub \& spokes} structure is a module $S$ with $t=1$.
\end{definition}

According to the above definitions, one can in fact determine the type of some group $S$ by considering Jaccard index~\cite{Jac01} of $S$ and $T$. We thus define a group type parameter $\tau$~\cite{SBB13}, $\tau\in\left[0,1\right]$, as
\begin{equation}
\tau(S,T)=\frac{|S\cap T|}{|S\cup T|}.
\label{equ:tau}
\end{equation}
Communities have $\tau=1$, whereas modules are indicated by $\tau=0$. Mixtures correspond to groups with $0<\tau<1$. For the remaining of the paper, we refer to groups with $\tau\approx 1$ or $\tau\approx 0$ as community-like and module-like groups, respectively.

The framework presented below is based on a group criterion $W$~\cite{SBB13}, $W\in\left[0,1\right]$.
\begin{equation}
W(S,T)=\mu(S,T)\left(1-\mu(S,T)\right)\left(\frac{L(S,T)}{st}-\frac{L(S,T^C)}{s(n-t)}\right),
\label{equ:W}
\end{equation}
where $L(S,T)$ is the number of links between $S$ and $T$, i.e., $L(S,T)=\sum_{\left(i,j\right)\in L}\delta(i\in S, j\in T)$, and $\mu(S,T)$ is the geometric mean of $s$ and $t$ normalized by the number of nodes $n$, $\mu\in\left[0,1\right]$.
\begin{equation}
\mu(S,T)=\frac{2s t}{n(s+t)}
\label{equ:mu}
\end{equation}

Notice that $W$ is an asymmetric criterion that favors the links between $S$ and $T$, and penalizes for the links between $S$ and $T^C$. Since the links with both endpoints in $S^C$ are not considered, $W$ is also a local criterion. We stress that, at least for the case $S=T$, criterion $W$ has a natural interpretation in a wide class of different generative graph models~\cite{ZLZ11a} (e.g., block models~\cite{WBB76}). Factor $\mu(1-\mu)$ in~\equref{W} prevents from extracting either very small or large groups with, e.g., $s=1$.

We next present the adopted group extraction framework~\cite{SBB13,ZLZ11a}. The framework extracts groups from the network sequentially, one by one, as follows. First, one finds group $S$ and its corresponding pattern $T$ that maximize criterion $W$ using, e.g., tabu search~\cite{Glo89} with varying initial conditions for $S$ and $T$. At each step of the search, a single node is swapped in either $S$ or $T$. Next, to extract the revealed group $S$ from the network, one removes merely the links between $S$ and $T$, and any node that might thus become isolated. The entire procedure is then repeated on the remaining network until criterion $W$ is larger than the value expected under the same framework in a corresponding Erd{\" o}s-R{\'e}nyi random graph~\cite{ER59}. The latter is estimated by a simulation, thus, all groups reported in the remaining of the paper are statistically significant at the $1\%$ level (see~\cite{ZLZ11a} for further details).

Note that the framework allows for overlapping~\cite{PDFV05}, hierarchical~\cite{RSMOB02}, nested and other classes of groups commonly found in real-world networks. Nevertheless, it explicitly guards against extracting groups that are not statistically significant. We refer to the network structure remaining after the extraction as \define{background}.

\begin{table}[!b]
\tbl{\label{tbl:groups}Node groups and corresponding patterns extracted from different networks.}
{\begin{tabular}{@{}cccccccc@{}} \toprule
Network & \multicolumn{3}{c}{Group} & Community & Core/periphery & Mixture & Module \\
 & \# & $\left<s\right>$ & $\left<t\right>$ & \multicolumn{4}{c}{\# ($\left<s\right>$)} \\ \colrule
\jbullet & $\hphantom{0}14$ & $\hphantom{0}9.0$ & $8.4$ & $\hphantom{00}5$ ($7.8$) & $1$ ($12.0$) & $\hphantom{0}6$ ($12.2$) & $\hphantom{0}2$ \hphantom{0}($5.5$) \\
\colt & $\hphantom{0}15$ & $10.3$ & $8.3$ & $\hphantom{00}3$ ($8.3$) & $1$ ($13.0$) & $\hphantom{0}9$ ($12.6$) & $\hphantom{0}2$ \hphantom{0}($6.5$) \\
\jung & $\hphantom{0}30$ & $\hphantom{0}8.7$ & $7.8$ & $\hphantom{0}18$ ($9.9$) & $1$ ($10.0$) & $\hphantom{0}5$ \hphantom{0}($9.6$) & $\hphantom{0}6$ \hphantom{0}($5.7$) \\
\lucene & $123$ & $12.1$ & $7.9$ & $\hphantom{0}55$ ($8.6$) & $2$ ($14.5$) & $27$ ($15.7$) & $39$ ($14.7$) \\ \colrule
\internet & $\hphantom{0}33$ & $10.6$ & $4.5$ & $\hphantom{00}1$ ($4.0$) & $1$ ($29.0$) & $\hphantom{0}3$ ($19.0$) & $28$ \hphantom{0}($9.6$) \\
\social & $160$ & $\hphantom{0}5.6$ & $5.6$ & $143$ ($5.6$) & $0$ \hphantom{0}($0.0$) & $12$ \hphantom{0}($6.8$) & $\hphantom{0}5$ \hphantom{0}($3.0$) \\ \botrule
\end{tabular}}
\begin{tabnote}
Networks are reduced to simple undirected graphs
\end{tabnote}
\end{table}

\subsection{\label{sec:groups:structure}Characteristic node group structure}
\tblref{groups} summarizes the basic properties of node groups extracted from different networks. Notice that the mean group size $\left<s\right>$ is somewhat comparable across software networks, where a characteristic group consists of around ten nodes. The mean pattern size $\left<t\right>$ is slightly smaller, but still comparable to $\left<s\right>$ (e.g., \jung network). On the other hand, $\left<s\right>\gg\left<t\right>$ for the Internet, due to an abundance of hub \& spokes-like modules. Since social networks are characterized by a pronounced community structure~\cite{NP03}, expectedly, $\left<s\right>\approx\left<t\right>$ for \social network.

\begin{figure}[!t]
	\centerline{\psfig{file=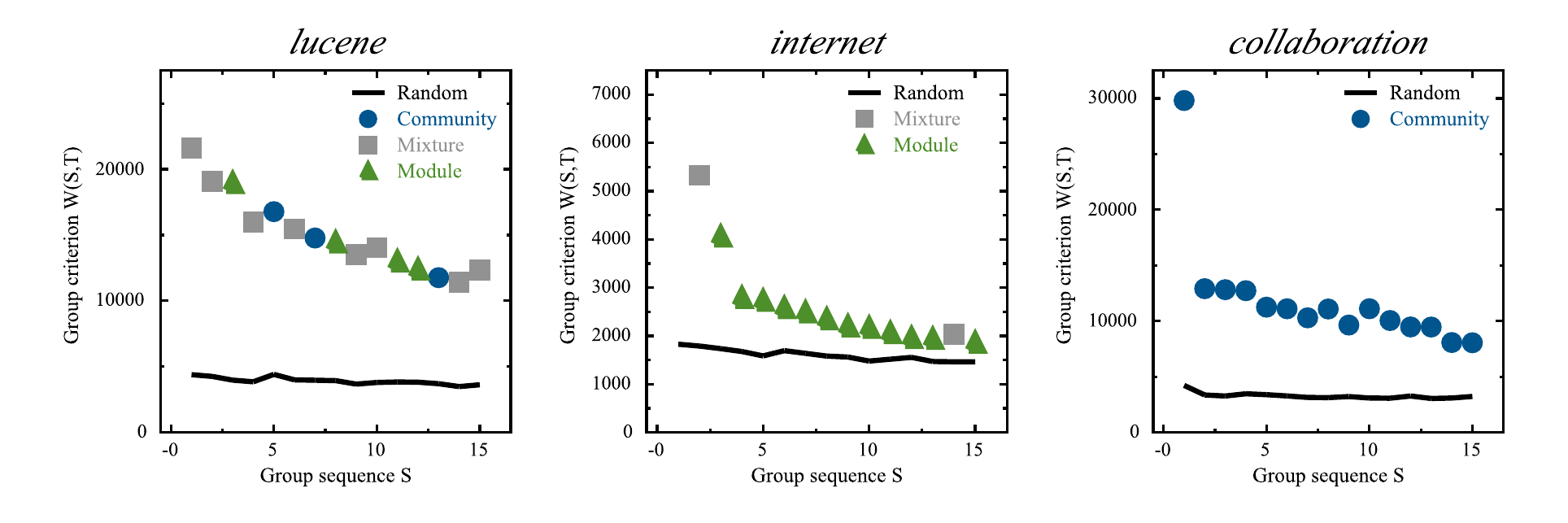, width=\plotswidth}}
	\caption{\label{fig:groups:criterion}Node group sequence extracted from larger networks (see also~\tblref{groups}). Note that \lucene software network contains communities, which are commonly found in social networks (e.g., \social network), modules like the Internet, and also different mixtures of these.}
\end{figure}

By examining the types of the revealed groups (see~\tblref{groups}), one observes a very clear distinction between different networks. As already indicated above, \social network consists of almost only communities. On the contrary, $85\%$ of the groups found in \internet network are modules. Software networks, however, are characterized by communities, modules and different mixtures of these (e.g., \lucene network). Thus, as already argued in the case of node mixing in~\secref{mixing}, software networks represent a unique mixture of dense community-like structure of social networks and sparse module-like topology of the Internet. For a better comprehension, \figref{groups:criterion} shows most significant groups extracted from the~networks.

\begin{table}[!b]
\tbl{\label{tbl:groups:structure}Node group structure revealed in different networks (see also~\tblref{groups}). Note that characteristic topology of different networks is well characterized by the mean group parameter $\left<\tau\right>$.}
{\begin{tabular}{@{}ccccccc@{}} \toprule
Network & Group & Community & Core/periph. & Mixture & Module & Background \\
 & $\left<\tau\right>$ & \multicolumn{5}{c}{\% Links (\% nodes)\tabmark{a}} \\ \colrule
\jbullet & $0.63$ & $15\%$ ($22\%$) & $\hphantom{0}8\%$ ($7\%$) & $53\%$ ($42\%$) & \hphantom{0}$6\%$ \hphantom{0}($7\%$) & $19\%$ ($66\%$) \\
\colt & $0.41$ & $\hphantom{0}7\%$ ($11\%$) & $\hphantom{0}5\%$ ($6\%$) & $69\%$ ($49\%$) & \hphantom{0}$4\%$ \hphantom{0}($6\%$) & $15\%$ ($64\%$) \\
\jung & $0.66$ & $62\%$ ($51\%$) & $\hphantom{0}3\%$ ($3\%$) & $12\%$ ($16\%$) & $10\%$ ($11\%$) & $12\%$ ($44\%$) \\
\lucene & $0.55$ & $19\%$ ($25\%$) & $\hphantom{0}1\%$ ($2\%$) & $30\%$ ($24\%$) & $38\%$ ($34\%$) & $11\%$ ($49\%$) \\ \colrule
\internet & $0.08$ & \hphantom{0}$0\%$ \hphantom{0}($1\%$) & $12\%$ ($4\%$) & $13\%$ \hphantom{0}($7\%$) & $34\%$ ($35\%$) & $41\%$ ($80\%$) \\
\social & $0.94$ & $71\%$ ($47\%$) & \hphantom{0}$0\%$ ($0\%$) & \hphantom{0}$6\%$ \hphantom{0}($5\%$) & \hphantom{0}$1\%$ \hphantom{0}($1\%$) & $22\%$ ($47\%$) \\ \botrule
\end{tabular}}
\begin{tabnote}
Networks are reduced to simple undirected graphs
\end{tabnote}
\begin{tabfootnote}
\tabmark{a} Nodes can be included in multiple overlapping groups
\end{tabfootnote}
\end{table}

Characteristic group structure of different networks is also reflected in the mean group parameter $\left<\tau\right>$ (\tblref{groups:structure}). Indeed, $\left<\tau\right>$ is almost zero or one for \internet and \social networks, respectively. For software networks, $\left<\tau\right>$ is between $0.4$ and $0.65$, as discussed above. \tblref{groups:structure} reports also the proportion of links explained by the group structure, and the proportion of nodes included in the groups. Despite the fact that group structure provides a rather coarse-grained abstraction of a network, the reveled groups explain $80$-$90\%$ of the links in software and social networks, and almost $60\%$ for the Internet. Also, groups contain most of the nodes in the networks.

As already discussed in~\secref{mixing:sickle}, different types of groups observed in software networks actually coincide with the intrinsic dynamics of the underlying software systems. More precisely, core classes of a software project commonly form dense inheritance hierarchies, while they also provide different convenience methods for transforming other core classes. Consequently, corresponding nodes in class dependency networks cluster together and form communities~\cite{SB11s,SB12u} (see~\figref{applications}). Moreover, software projects commonly consist of classes that represent independent implementations of the same functionality (e.g., different group detection algorithms). By definition, these do not depend on each other; however, they do depend on a similar set of other classes. Hence, corresponding nodes in software networks aggregate together into module-like groups~\cite{SB12u,SB12s} (see~\figref{applications}). Similarly as above, mixtures of nodes in software networks are often just an artifact of different programming principles and practical limitations of software systems.

Notice also particularly module-like structure of \colt network compared to other software networks (see $\left<\tau\right>$ in~\tblref{groups:structure}). Since the network represents a software library for complex scientific and technical computing, high performance and scalability are of much greater importance than the system extensibility and future reusability. While the latter implies a modular design according to minimum-coupling and maximum-cohesion paradigm~\cite{SMC99} and, consequently, a community-like structure of software networks~\cite{SB11s}, the former demands a great deal of code duplication, which in fact promotes module-like groups in software networks~\cite{SB12u}. Equivalently, networks that correspond to software projects with particularly modular design reveal more community-like structure (e.g., \jung network). Group structure of software networks thus reflects different programming principles and paradigms followed during project development, which could be used for software quality control.

Preliminary work on practical applications of network group detection in software engineering is described in~\secref{applications}, while, in the following section, we relate the characteristic group structure of software networks to previously observed dichotomous node degree mixing and degree-corrected clustering assortativity.

\subsection{\label{sec:groups:mixing}Group degree and clustering mixing}
\secref{mixing} shows that software networks are characterized by dichotomous node degree mixing that is assortative from the perspective of out-degrees, and disassortative from the perspective of in-degrees. Moreover, networks are composed of regions with rather similar clustering and reveal strong degree-corrected clustering assortativity. We have postulated a hypothesis that the observed structure is a consequence of different types of groups of nodes present in the networks. More precisely, software networks contain dense community-like groups in regions with higher clustering, which imply assortativity in the out-degree, and sparse module-like groups in regions with lower clustering, which promote disassortativity by in-degree, and different mixtures of these. As already discussed before, existence of different groups immediately explains also degree-corrected clustering assortativity.

\begin{figure}[!t]
	\centerline{\psfig{file=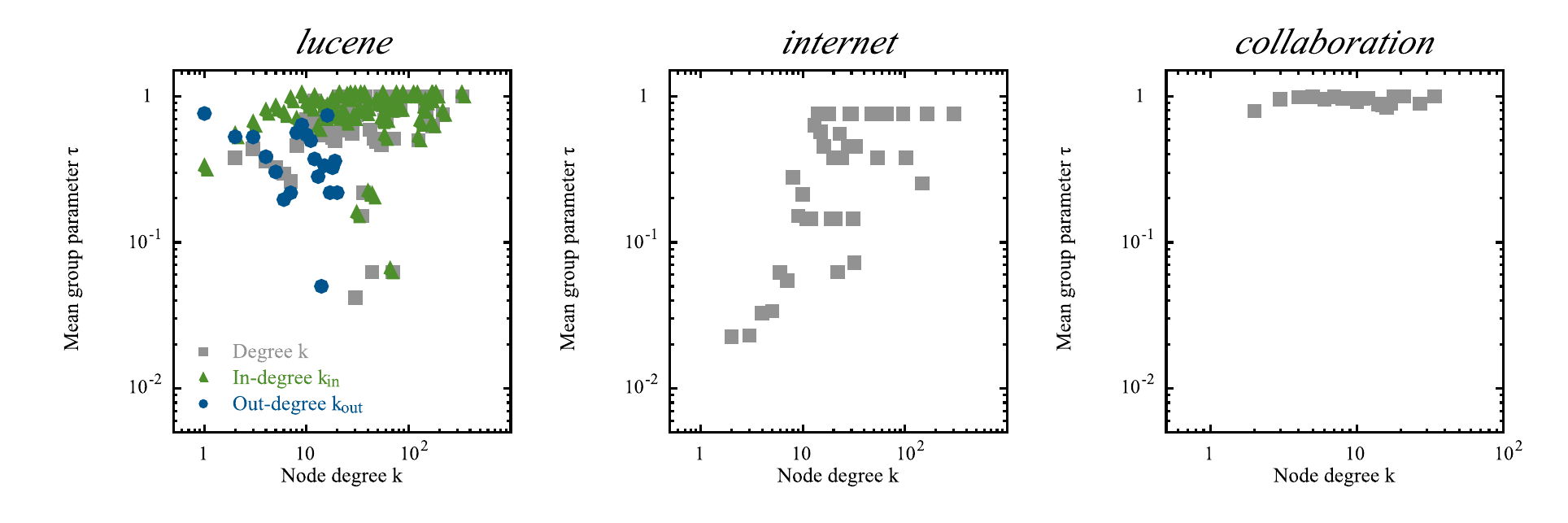, width=\logplotswidth}}
	\caption{\label{fig:groups:degree:profile}Group degree profiles of larger networks that reveal no characteristic scaling.}
\end{figure}

\begin{figure}[!b]
	\centerline{\psfig{file=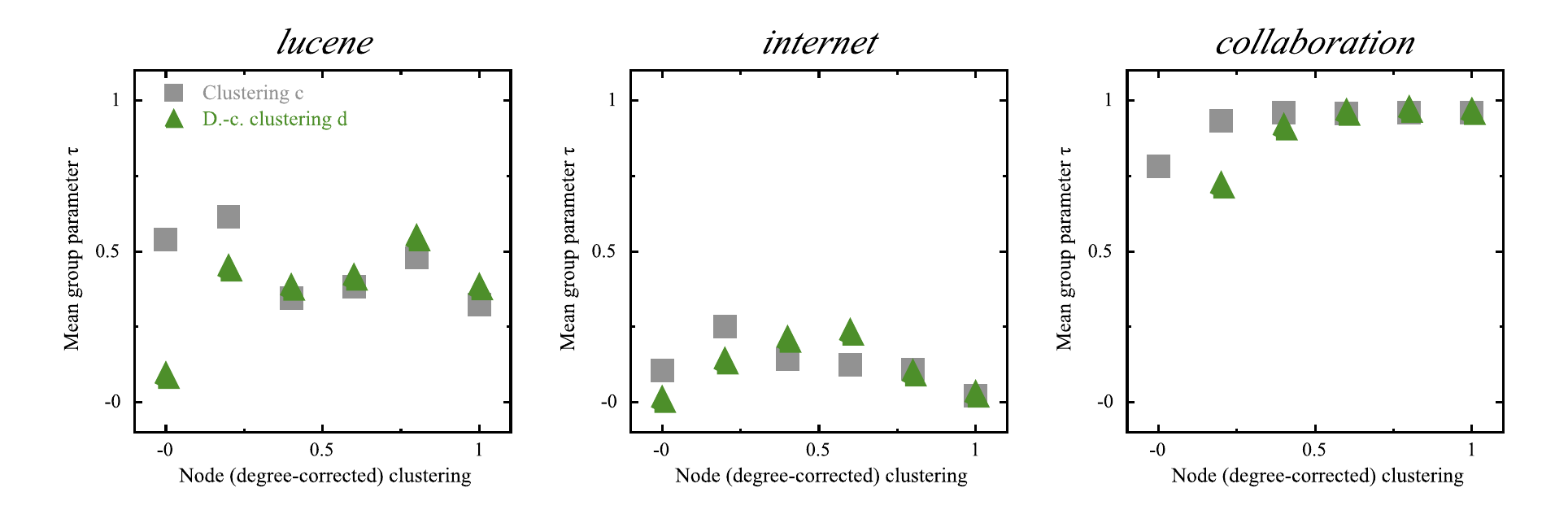, width=\plotswidth}}
	\caption{\label{fig:groups:clustering:profile}Group (degree-corrected) clustering profiles of larger networks. Note that \lucene software network consists of module-like groups with $\tau\approx 0$ in regions with $d\approx 0$ as the Internet and mostly community-like groups with $\tau\approx 1$ in regions with $d\approx 1$ as the \social network.}
\end{figure}

We pursue the hypothesis by first investigating the regions of the networks occupied by different types of groups. \figref{groups:degree:profile} shows group degree profiles that plot mean group parameter $\left<\tau\right>$ against node degree $k$. These do not provide any clear insight into the structure of the networks, due to a rather extensive overlaps between the groups, i.e., both high and low degree nodes are included into different groups. On the other hand, group degree-corrected clustering profiles in~\figref{groups:clustering:profile} clearly show that software network indeed consists of module-like groups with $\tau\approx 0$ in sparse regions with low clustering $d\approx 0$ as hypothesized, while the plot reveals an expected increasing trend. Similarly, the network contains mostly community-like groups with $\tau\approx 1$ in dense regions with high clustering $d\approx 1$; however, the corresponding nodes are included also in overlapping module-like groups thus $\tau\approx 0.5$ (see~\figref{groups:clustering:profile}). The same observations apply for social network and the~Internet.

We next consider group degree and clustering mixing. For this purpose, we define group degree mixing coefficient $\tilde{r}$, $\tilde{r}\in\left[-1,1\right]$, as
\begin{equation}
\tilde{r}=\frac{1}{\sigma_{\tilde{k}_S}\sigma_{\tilde{k}_T}}\sum_{S,T}\left(\tilde{k}_S-\left<\tilde{k}_S\right>\right)\left(\tilde{k}_T-\left<\tilde{k}_T\right>\right),
\label{equ:r:tilde}
\end{equation}
where $\tilde{k}_S$ is the degree of group $S$, i.e., $\tilde{k}_S=\sum_{i\in S}k_i/s$, and similarly for the pattern degree $\tilde{k}_T$. We further define also directed group degree mixing
coefficients $\tilde{r}_{(\alpha,\beta)}$, $\alpha,\beta\in\left\{in,out\right\}$,
 and group clustering mixing coefficients $\tilde{r}_c$ and $\tilde{r}_d$, symmetrically as in~\secref{mixing}. These provide an overview of degree and clustering mixing in regions covered by groups of nodes, and enable reasoning about the network structure implied by different~types~of~groups.

\begin{figure}[!t]
	\centerline{\psfig{file=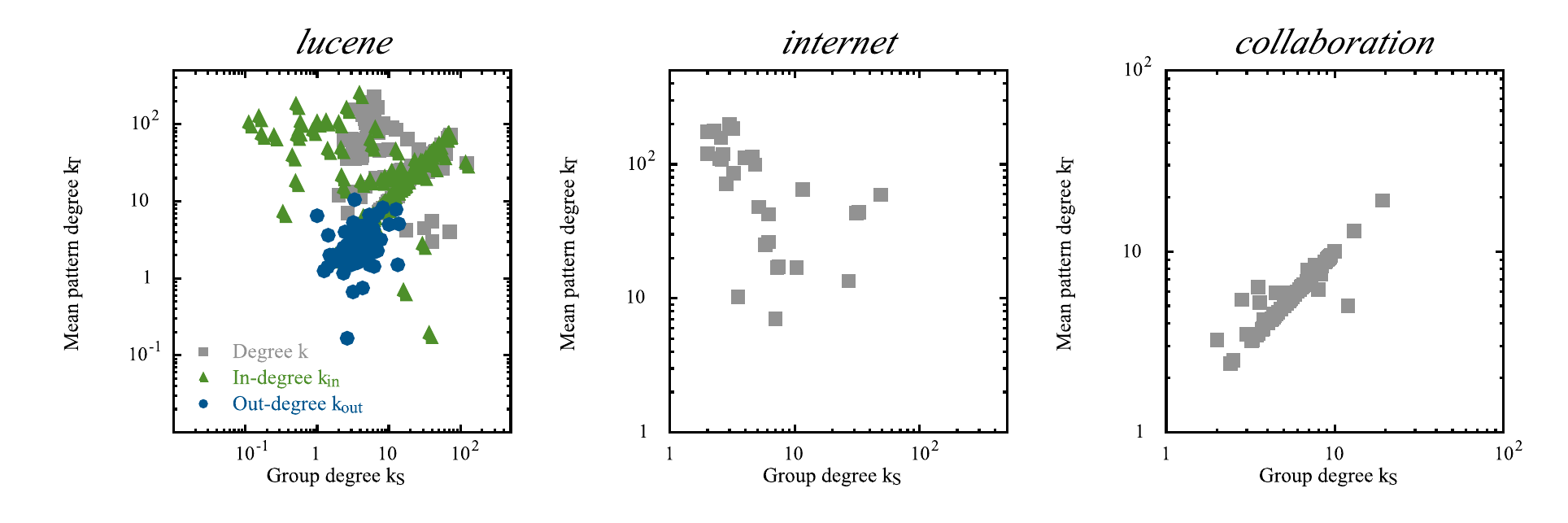, width=\logplotswidth}}
	\caption{\label{fig:groups:degree:mixing}Group pattern connectivity plots of larger networks (see also~\tblref{groups:mixing}). Note that \lucene software network reveals assortative mixing by out-degree as social networks (e.g., \social network) and disassortative mixing by in-degree as the Internet. While the former is an artifact of community-like groups, the latter is in fact a signature module-like groups.}
\end{figure}

\tblref{groups:mixing} displays group mixing coefficients. Most evidently, almost all correlations observed in the case of node mixing are strictly enhanced (see~\tblref{mixing:dichotomous}). Social network is assortative by degree, while the Internet is degree disassortative. Software networks again reveal disassortativity in the in-degrees. However, in contrast to before, group structure in fact promotes assortativity by out-degree in all software networks except \colt network, due to the reason given in~\secref{groups:structure}. \figref{groups:degree:mixing} shows also group pattern connectivity plots. For software network, one can clearly observe an increasing trend in the case out-degrees, and also larger in-degrees, which is obviously an artifact of community-like groups, as in the case of social network. Otherwise, in-degree profile has a decreasing structure similar to that of the Internet, which signifies module-like groups. Thus, confirming the above hypothesis, group structure of software networks can indeed explain dichotomous degree mixing with module-like groups responsible for disassortativity, most notably seen for smaller in-degrees, and community-like groups promoting assortativity in the out-degrees.

\begin{table}[!h]
\tbl{\label{tbl:groups:mixing}Group degree and clustering mixing coefficients of different networks.}
{\begin{tabular}{@{}cccccccc@{}} \toprule
Network & $\tilde{r}$ & $\tilde{r}_{(in,in)}$ & $\tilde{r}_{(in,out)}$ & $\tilde{r}_{(out,in)}$ & $\tilde{r}_{(out,out)}$ & $\tilde{r}_c$ & $\tilde{r}_d$ \\ \colrule
\jbullet & $-0.02$ & $-0.15$ & $-0.01$ & $-0.20$ & $\hphantom{-}0.66$ & $\hphantom{-}0.47$ & $\hphantom{-}0.97$ \\
\colt & $-0.63$ & $-0.60$ & $-0.27$ & $-0.63$ & $-0.17$ & $-0.59$ & $\hphantom{-}0.76$ \\
\jung & $-0.32$ & $-0.32$ & $-0.12$ & $-0.30$ & $\hphantom{-}0.54$ & $\hphantom{-}0.45$ & $\hphantom{-}0.78$ \\
\lucene & $-0.16$ & $-0.19$ & $-0.12$ & $-0.22$ & $\hphantom{-}0.39$ & $\hphantom{-}0.17$ & $\hphantom{-}0.85$ \\ \colrule
\internet & $-0.54$ & - & - & - & - & $-0.37$ & $\hphantom{-}0.37$ \\
\social & $\hphantom{-}0.84$ & - & - & - & - & $\hphantom{-}0.81$ & $\hphantom{-}0.95$ \\ \botrule
\end{tabular}}
\end{table}

It ought to be mentioned that the above relation between degree mixing and different groups of nodes can be justified theoretically. Since $S=T$ for communities, this implies degree assortativity, as long as the sizes of communities differ~\cite{NP03}. Also, for $s\not\approx t$, module-like groups should result in degree disassortativity~\cite{SB12u}. Finally, according to discussion in~\secref{groups:structure}, modules or communities are best pronounced through the out-degrees and in-degrees of nodes, respectively.

\begin{figure}[!b]
	\centerline{\psfig{file=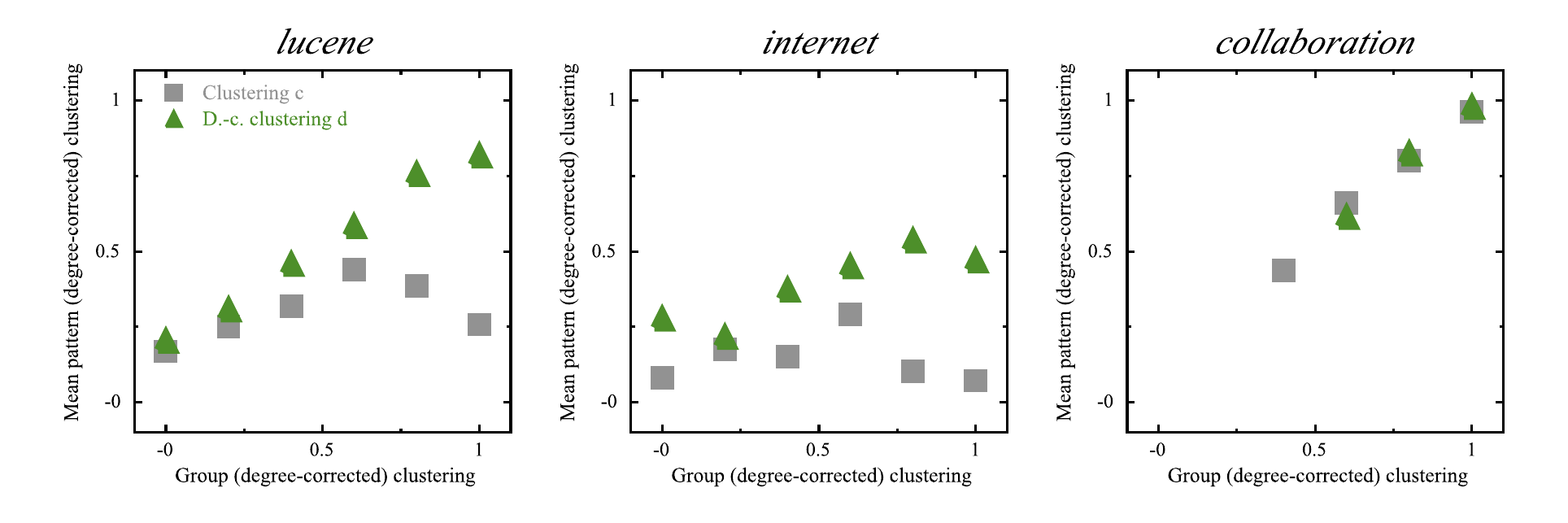, width=\plotswidth}}
	\caption{\label{fig:groups:clustering:mixing}Group pattern (degree-corrected) clustering plots of larger networks (see also~\tblref{groups:mixing}). Note~that networks reveal extremely clear group degree-corrected clustering~\cite{SV05} assortativity (e.g., \lucene and \social network), which is an indication of a well pronounced group structure.}
\end{figure}

\tblref{groups:mixing} also reports group clustering mixing coefficients. As before, $\tilde{r}_c<0$ in some degree disassortative networks, due to the biases introduced in clustering $c$ (see~\secref{mixing:sickle}). Nevertheless, degree-corrected clustering mixing $\tilde{r}_d$ signifies extremely assortative structure with correlations between $0.75$ and $0.95$ for software and social networks (see also \figref{groups:clustering:mixing}). Presence of clear groups of nodes thus indeed implies degree-corrected clustering assortativity, while the value of $\tilde{r}_d$ can be related to the quality of network group structure. For example, in the case of the Internet, which has least clear group structure (see~\secref{groups:structure}), $\tilde{r}_d$ is only $0.37$.

In summary, characteristic groups of nodes provide an important insight into the dynamics of complex networks and can, at least to some extent, explain the unique structure of software networks (i.e., degree and clustering mixing). There is of course no reason why the same principles should not apply to other real-world networks, directed or undirected, which will be thoroughly explored in future work.

\section{\label{sec:applications}Applications in software engineering}
\begin{figure}[!t]
\centerline{
\subfigure[\label{fig:applications:jung}Community in \jung network ($\tau=1$)]{
\psfig{file=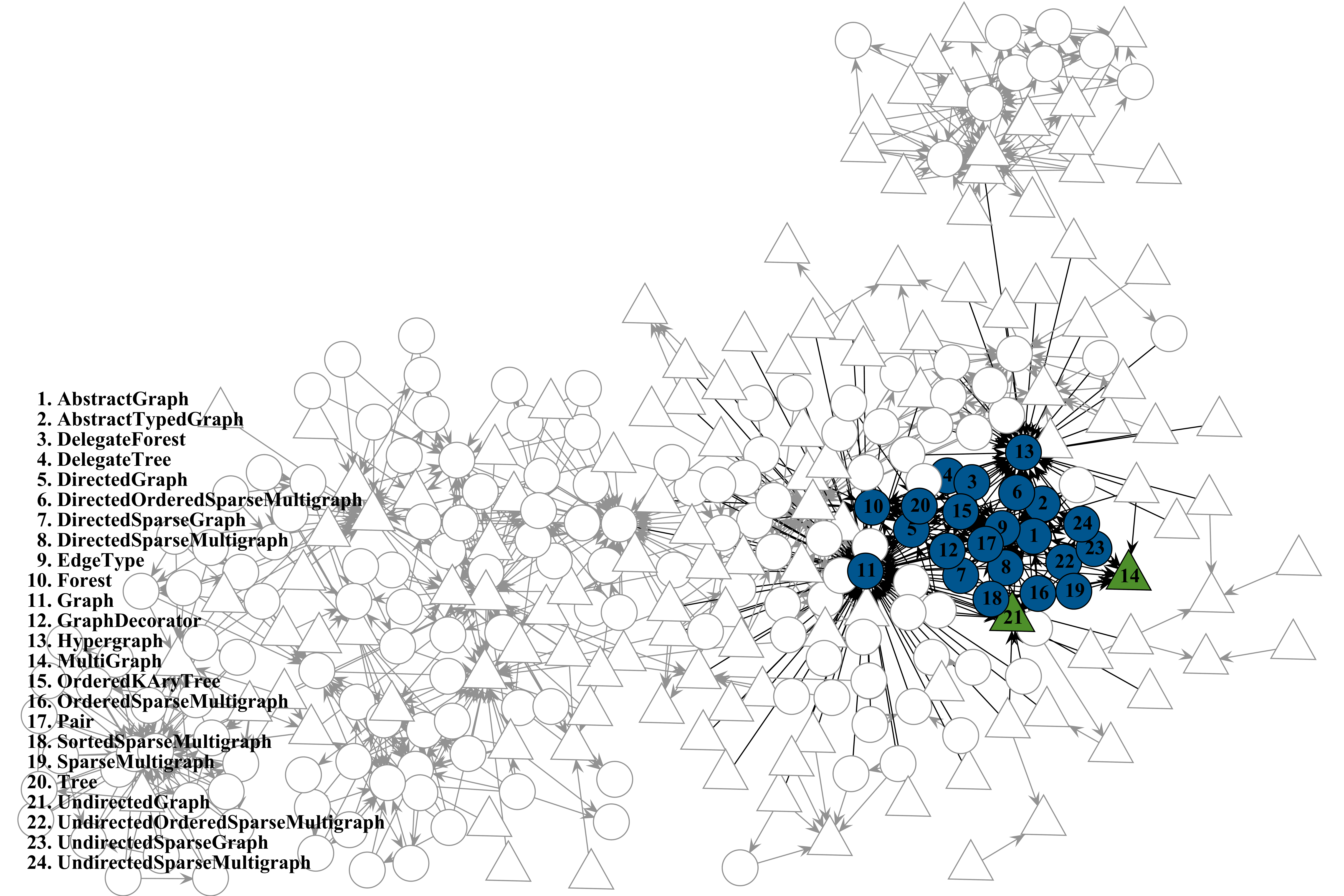, width=\subfigswidth}}
\subfigure[\label{fig:applications:colt}Module-like group in \colt network ($\tau=0.06$)]{
\psfig{file=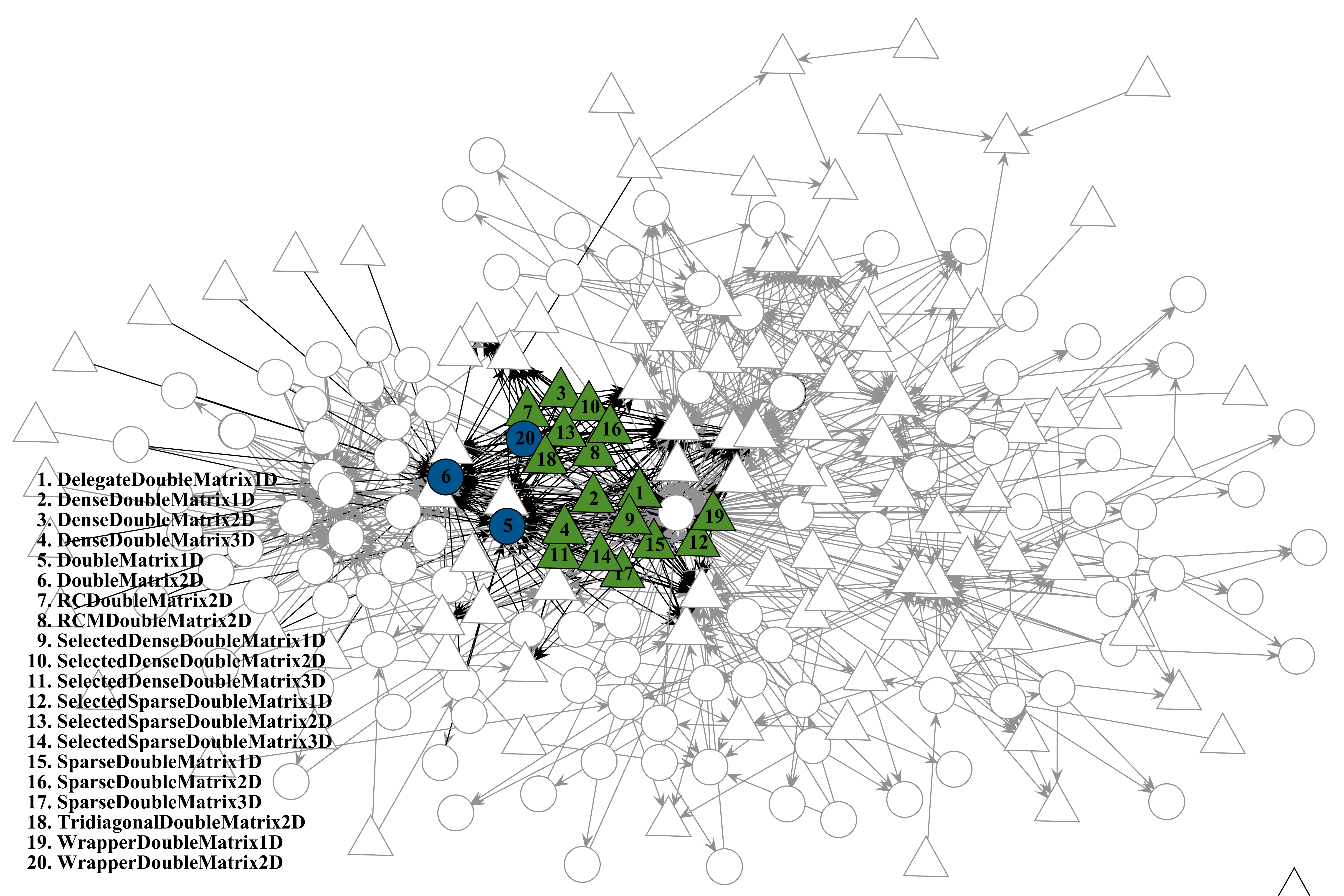, width=\subfigswidth}}}
	\caption{\label{fig:applications}Most significant groups of nodes extracted from different software networks (see also~\tblref{groups}). The groups correspond to (a)~core classes of the software project and (b)~different implementations of classes with the same functionality. (Nodes with degree-corrected clustering~\cite{SV05} above or below the mean are shown as circles and triangles, respectively.)}
\end{figure}

The present section describes preliminary work on practical applications of network group detection in software engineering. As already discussed before, groups of nodes in software dependency networks coincide with the intrinsic properties of the underlying software systems. For instance, \figref{applications} shows the most significant groups revealed in \jung and \colt networks. In the case of the former, the best group is a community that corresponds to core classes of the project, as predicted in~\secref{groups:structure}. Since the network represents a framework for graph and network analysis, these are actually different graphs, multigraphs, hypergraphs and trees. Notice that the revealed group is not only very clear, but also rather exhaustive.

On the other hand, the most significant group in \colt network, which represents a software library for high-performance scientific computing, is module-like and contains different implementations of matrices (e.g., dense, sparse or wrapped). Recall that the latter is consistent with the rationale behind the existence of modules in software networks given in~\secref{groups:structure}. Similarly as above, the group is indeed transparent, while the identifiers of the corresponding software classes are extremely consistent with each other (see~\figref{applications:colt}). Thus, one can in fact derive templates for class identifiers (e.g., by mining common textual patterns~\cite{BR99}) and unique class dependencies on the level of groups of nodes in a software network (i.e., by analyzing corresponding node patterns). These can be adopted in future project development, in order to maintain a high consistency of a software system, to reduce code duplication issues and other. Furthermore, one can also predict the package of a~class.

Classes of object-oriented software systems are organized into software packages that form a complex hierarchy. Each class is a member of exactly one package, whereas the classes can reside also in the inner nodes of the package hierarchy. For example, the group of nodes shown in~\figref{applications:jung} consists mostly of classes in \texttt{edu.uci.ics.jung.graph} package, while the group in~\figref{applications:colt} represents classes in \texttt{cern.colt.matrix.impl} package. To predict the package of some class given the group structure of the software network, we investigate the classes, whose nodes are residing in the same network groups as the concerned one. These classes are then weighted according to the Jaccard similarity~\cite{Jac01} between the corresponding nodes' neighborhoods and their packages are taken as the candidates for the prediction. We select the most frequent package with respect to weights, while ties are broken uniformly at random (see~\cite{SB11s,SB12s} for details). Note that, instead of considering nodes within the same network groups, one can of course examine merely nodes' neighbors or the entire network. For comparison, we also report the performance of a classifier that predicts the most frequent (i.e., majority) package within the software system for each class and a random classifier. However, the adoption of some more sophisticated approaches like deep belief nets~\cite{HOT06} or structured support vector machine~\cite{TJHA05} would inevitably require the identification of learning features.

\tblref{applications:prediction:package} shows classification accuracy for software package prediction. Observe that the accuracy for the strategy based on network groups is around $75\%$ in all cases except for the larger \lucene network. We stress that the latter is an impressive result. Indeed, the task at hand represents an extremely difficult classification problem due to a large number of possible classifications, while this number is else two or three in most practical applications (see performance of the baseline classifiers). Notice also that the strategy based on nodes' neighbors performs very well in \jbullet and \jung networks with more community-like groups (see $\left<\tau\right>$ in~\tblref{groups:structure}), since the groups well coincide with nodes' neighborhoods. On the other hand, the neighbors are in fact different from one another in \colt network with more module-like groups (see~\secref{groups}), which significantly decreases the performance.

\begin{table}[!h]
\tbl{\label{tbl:applications:prediction:package}Classification accuracy of software package prediction based on the node's neighbors~$\Gamma$ or groups $S$, or the entire network $N$ (see text for details).}
{\begin{tabular}{@{}cccccccc@{}} \toprule
Network & \# Classes\tabmark{a} & \# Packages & $\Gamma$ & $S$ & $N$ & {\itshape Majority} & {\itshape Random} \\ \colrule
\jbullet & $\hphantom{0}107$ & $\hphantom{0}11$ & $72.0\%$ & $\mathbf{75.7\%}$ & $64.5\%$ & $\mathit{28.0\%}$ & $\mathit{8.6\%}$ \\
\colt & $\hphantom{0}154$ & $\hphantom{0}16$ & $58.4\%$ & $\mathbf{73.4\%}$ & $55.2\%$ & $\mathit{22.7\%}$ & $\mathit{5.9\%}$ \\
\jung & $\hphantom{0}237$ & $\hphantom{0}31$ & $72.2\%$ & $\mathbf{74.2\%}$ & $65.0\%$ & $\mathit{11.4\%}$ & $\mathit{3.3\%}$ \\
\lucene & $1335$ & $178$ & $47.1\%$ & $\mathbf{49.2\%}$ & $43.7\%$ & $\mathit{\hphantom{0}6.4\%}$ & $\mathit{0.6\%}$ \\
\botrule
\end{tabular}}
\begin{tabnote}
Results are averages over $100$ runs
\end{tabnote}
\begin{tabfootnote}
\tabmark{a} Analysis is reduced to nodes included in network groups
\end{tabfootnote}
\end{table}

\tblref{applications:prediction:highlevel} shows also the accuracy for high-level software package prediction problem, where we consider only the packages at the topmost level of the package hierarchy. For \jung network, these are \texttt{graph}, \texttt{algorithms}, \texttt{io}, \texttt{visualization} and \texttt{visualization3d} (prefix \texttt{edu.uci.ics.jung} is omitted). Again, the strategy based on network groups performs particularly well with classification accuracy around $85$-$90\%$. Besides, the strategy based on nodes' neighbors, and also the network-based strategy for \jung network, obtains surprisingly high results, which further justifies the construction of software dependency networks (see~\secref{networks}).

\begin{table}[!t]
\tbl{\label{tbl:applications:prediction:highlevel}Classification accuracy of high-level software package prediction based on the node's neighbors~$\Gamma$ or groups $S$, or the entire network $N$ (see text for details).}
{\begin{tabular}{@{}cccccccc@{}} \toprule
Network & \# Classes\tabmark{a} & \# Packages & $\Gamma$ & $S$ & $N$ & {\itshape Majority} & {\itshape Random} \\ \colrule
\jbullet & $\hphantom{0}107$ & $\hphantom{0}5$ & $\mathbf{84.6\%}$ & $\mathbf{85.0\%}$ & $78.5\%$ & $\mathit{64.5\%}$ & $\mathit{20.4\%}$ \\
\colt & $\hphantom{0}154$ & $10$ & $\mathbf{86.4\%}$ & $83.8\%$ & $69.5\%$ & $\mathit{39.0\%}$ & $\mathit{\hphantom{0}9.7\%}$ \\
\jung & $\hphantom{0}237$ & $\hphantom{0}5$ & $89.1\%$ & $\mathbf{90.5\%}$ & $\mathbf{91.1\%}$ & $\mathit{44.3\%}$ & $\mathit{20.3\%}$ \\
\lucene & $1335$ & $15$ & $85.5\%$ & $\mathbf{90.8\%}$ & $85.0\%$ & $\mathit{28.2\%}$ & $\mathit{\hphantom{0}6.6\%}$ \\
 \botrule
\end{tabular}}
\begin{tabnote}
Results are averages over $100$ runs
\end{tabnote}
\begin{tabfootnote}
\tabmark{a} Analysis is reduced to nodes included in network groups
\end{tabfootnote}
\end{table}

Thus, characteristic group structure of software networks can indeed be exploited to quite accurately infer the package hierarchy of software systems~\cite{SB11s,SB12s}. This has numerous applications. For instance, the framework can be used to predict packages of new classes introduced into an unknown software project or even the programming language itself, to detect possibly duplicated classes, or for merging classes across different software packages or libraries (one by one). Such tasks would else demand significant manual labor, especially for large and complex software systems. Furthermore, network group detection can be adopted for software project refactoring, in order to derive either more modular or more functional software package hierarchy~\cite{SB12s,SB12u} (i.e., community-like and module-like, respectively).

As shown below, characteristic groups in software networks can also be used to infer the name of the developer that implemented a particular class, the exact version at which it was introduced into the project or its type (i.e., class or interface). However, as this information was largely unavailable or could not be obtained automatically for the software projects considered, we only report the results for \jung network. The prediction else proceeds exactly the same as before, while the classes with unknown version or author information are grouped into a single category.

\begin{table}[!b]
\tbl{\label{tbl:applications:prediction:other}Classification accuracy of class prediction for \jung software network based on the node's neighbors~$\Gamma$ or groups $S$, or the entire network $N$ (see text for details).}
{\begin{tabular}{@{}ccccccc@{}} \toprule
Prediction & \# Categories & $\Gamma$ & $S$ & $N$ & {\itshape Majority} & {\itshape Random} \\ \colrule
Class type & $\hphantom{0}2$ & $65.0\%$ & $\mathbf{85.2\%}$ & $\mathbf{84.8\%}$ & $\mathit{\mathbf{84.4\%}}$ & $\mathit{49.9\%}$ \\
Class version & $\hphantom{0}9$ & $67.7\%$ & $\mathbf{72.8\%}$ & $66.2\%$ & $\mathit{44.3\%}$ & $\mathit{11.2\%}$ \\
Class author & $11$ & $\mathbf{71.6\%}$ & $\mathbf{71.0\%}$ & $\mathbf{70.9\%}$ & $\mathit{44.3\%}$ & $\mathit{\hphantom{0}9.2\%}$ \\
 \botrule
\end{tabular}}
\begin{tabnote}
Results are averages over $100$ runs
\end{tabnote}
\end{table}

\tblref{applications:prediction:other} shows the classification accuracy for different software prediction problems. For class type prediction, the strategy based on network groups performs only slightly better than the baseline approach that classifies all software classes into the same category. On the other hand, the performance is significantly improved in the case of class version and author prediction problems with accuracy over $70\%$. This is not very surprising, since classes with the same functionality that appear as different groups in software networks are commonly introduced within the same version of the software project and implemented by the same developer.

Furthermore, according to~\secref{groups:structure}, the quality of network group structure reflects different programming principles and paradigms. Since this can be measured by degree-corrected group clustering mixing (see~\secref{groups:mixing}), the latter enables different applications in software development and quality control.

\section{\label{sec:conclusions}Conclusions and future work}
The present paper rigorously analyzes the structure of complex software networks. These can be seen as an interplay between a dense structure of social networks and a sparse topology of the Internet. In particular, we show that software networks reveal characteristic node group structure, which consists of dense communities, sparse module-like groups and also different mixtures of these. Communities imply assortative mixing by degree, whereas just the opposite holds for the modules. Thus, software networks reveal dichotomous degree mixing that is assortative in the out-degrees and disassortative in the in-degrees. Furthermore, communities appear in denser regions with higher clustering, while most pronounced modules occupy sparse regions with very low clustering. The latter in fact promotes degree-corrected clustering assortativity, which is observed in all of the networks analyzed.

Besides, the group structure of software networks also coincides with the intrinsic properties of the underlying software systems. The paper thus includes some preliminary work on practical applications of network group detection in software engineering. Nevertheless, their true practical value in real scenarios remains somewhat unclear and will be more throughly investigated in the future.


The study of differences between software and social networks, and the Internet,
reveals notably distinct network topologies that are most likely governed by different phenomena.
We stress that dichotomous node degree mixing has not yet been observed in the case of directed networks.
Furthermore, preliminary results show that the existing graph models do not produce 
degree-corrected clustering assortativity of real-world networks. The latter
will be the main focus of our future~work.

Additionally, 
the paper implies several other prominent directions for future research. First, 
the observed node mixing and group structure might also apply to different software and other real-world networks. Among these, various information networks seem most promising. Next, characteristic group structure revealed for software networks might be further related to other properties, e.g., self-similarity~\cite{BSB12} or hierarchical structure~\cite{VS07}. Last, although we provide some rationale for the presence of groups in software networks, a generative graph model is still an open~issue.


\section*{Acknowledgments}
This work has been supported in part by the Slovenian Research Agency Program No.~P2-0359, by the Slovenian Ministry of Education, Science and Sport Grant No. 430-168/2013/91, and by the European Union, European Social Fund.

\appendix


\end{document}